\documentclass{optica-article}

\journal{opticajournal} % for journals or Optica Open

\articletype{Research Article}

\usepackage{lineno}
\usepackage{amsmath} % Make sure to include this package
\usepackage{graphicx}% Include figure files
\usepackage{dcolumn}% Align table columns on decimal point
\usepackage{bm}% bold math
\usepackage{subcaption} % Include in your preamble
\usepackage{tikz}
\usepackage{pgfplots}
\pgfplotsset{compat=newest}
\usetikzlibrary{patterns}
\usetikzlibrary{decorations.pathmorphing,shapes}
\usetikzlibrary{arrows.meta} % This line is important
\usepgfplotslibrary{fillbetween} % For filling between lines
\usetikzlibrary{arrows.meta, bending, decorations.markings}
\usetikzlibrary{arrows,decorations.markings,shapes.geometric,calc,intersections}
\usetikzlibrary{3d}
\usepackage{asypictureB}
\usepackage{asymptote} % Make sure to include this package in your preamble
\usepackage{placeins} % Ensure this is in your preamble
\usepackage{enumitem}
\usepackage{comment}
\begin{document}

\title{Analysis of Bessel Beam Generation Using MetaMaterials in Photonic Integrated Circuits}

\author{Solomon Serunjogi,\authormark{1,*} and Mahmoud S. Rasras,\authormark{1}}

\address{\authormark{1} Department of Electrical and Computer Engineering, New York University Abu Dhabi, P.O. Box 129188, Abu Dhabi, United Arab Emirates.\\ }

\email{\authormark{*}sms10215@nyu.edu, mr5098@nyu.edu} %% email address is required; see note below about the corresponding author designation

% use {asbstract*} to suppress the copyright line. Copyright information will be added in production

\begin{abstract*} 
 
Bessel beams, known for their unique non-diffracting property, maintain their shape and intensity over long distances, making them invaluable for applications in optical trapping, imaging, and communications. This work presents a comprehensive theoretical analysis of micro-photonic antennas designed to generate Bessel beams within the Terahertz (THz) and optical frequency ranges. The technique is demonstrated by generating far-field patterns of Bessel waves at these frequencies. The design employs metasurface patterns arranged as arrays of concentric rings atop rectangular silicon waveguides, collectively creating a Bessel beam. Dyadic Green's function integral equation techniques are used to model the transverse electric (TE) and transverse magnetic (TM) fields in the metasurface radiation zone. Utilizing orthogonal vector wave functions, Bloch theorem, Floquet harmonics, and the transverse resonance technique, a photonic chip is designed to achieve a non-diffracting range of 500 $\mu$m at the optical telecom wavelength of 1.5 $\mu$m for a metasurface radius of 25 $\mu$m. A radiation efficiency exceeding 80$\%$ is achieved by optimizing the attenuation constant ($\alpha$) along the structure. The theoretical models are validated through simulations for both optical (1.5$\mu$m) and Terahertz (14 $\mu$m) wavelengths, demonstrating significant alignment between predictions and simulation results.

\end{abstract*}

%%%%%%%%%%%%%%%%%%%%%%%%%%  body  %%%%%%%%%%%%%%%%%%%%%%%%%%
\section{Introduction}
The study of Bessel beams has attracted significant interest in many areas of engineering and physical sciences due to their diffraction-free nature and self-healing properties \cite{monticone2015leaky,khonina2020bessel,vicente2021bessel}. These properties render Bessel beams highly advantageous in integrated photonics, particularly within communication systems, where they contribute to extended propagation ranges, non-diffracting characteristics, improved imaging capabilities, and efficient handling of optical signals \cite{mphuthi2019free}.  Furthermore, the non-diffracting nature of Bessel beams renders them particularly valuable for optical trapping \cite{ettorre2018near,vcivzmar2006sub} useful in biological and imaging applications\cite{an2021direct,vcivzmar2006sub}.

Bessel beams are predominantly generated in free-space environments or within systems exhibiting cylindrical symmetry, such as cylindrical waveguide cavities at microwave frequencies \cite{ettorre2012generation,pakovic2021bessel}. This generation often employs radiation slots on metallic surfaces to produce leaky modes from the cavity into free space \cite{ettorre2012generation,luukkonen2008simple}. However, in the context of rectangular waveguides, such as those made of silicon for THz and optical frequencies, Bessel solutions do not naturally occur as modes. These waveguides primarily support guided modes in harmonic sines and cosines, which differ markedly from the nature of Bessel beams \cite{collin1990field}. Therefore, to introduce a Bessel-like beam into a rectangular waveguide, it is imperative to integrate inhomogeneous metasurfaces \cite{engheta2006metamaterials} to project the k-vectors onto a cone. The interaction between an incident beam and the metasurface lattice leads to scattering and diffraction, thereby modifying the beam's phase and amplitude as required. 

This paper presents a theoretical model utilizing vector wave functions and Dyadic Green's functions to analyze a cylindrical metasurface composed of concentric silicon rings. The analysis is supported by the Bloch Theorem for periodic cylindrical structures and the transverse resonance technique (TRT) for impedance synthesis. Simulations are performed using CST Microwave Studio \cite{studio2008cst} to validate the derived models. The study concludes with a summary of the key findings.

 %The paper is organized as follows: Section I provides an overview of early efforts in generating Bessel beams in free space. Section II delves into the mathematical modeling of Green's functions as a tool for describing propagation in bounded homogeneous silicon waveguides. In Section III, we present an analysis of inhomogeneous metasurfaces using Bloch analysis and the transverse resonance method for impedance synthesis. Subsequently, Section IV presents simulation results of concentric cylindrical rings atop a silicon waveguide, validating the models discussed in Sections II and III. Finally, Section V offers a conclusion summarizing the key findings and implications of this study.

%%Adherence to the specifications listed in this template is essential for efficient review and publication of submissions. Proper reference format is especially important (see Section \ref{sec:refs}). For submissions to the Optica Open preprint server, many of the style and format guidelines described in this template are not applicable, although adhering to the instructions in this template will ease the process of converting a preprint to an Optica Publishing Group journal submission. You may find it helpful to use our optional \href{https://preflight.paperpal.com/partner/optica/opticapublishinggroupjournals}{Paperpal} manuscript readiness check and \href{https://languageediting.optica.org/}{language polishing service}.

\section{Bessel beams in free space}\label{BB_fs}

In 1987, Durnin \cite{durnin1987exact} made a notable discovery by identifying a unique set of solutions to the scalar wave equation in cylindrical coordinates. It was noted that the Bessel function is a solution of the Helmholtz wave equation, and its mathematical properties depict propagating waves with non-diffracting characteristics. This phenomenon gives rise to a class of waves known as "nondiffracting" or diffraction-free wave fields. In simpler terms, these wave fields maintain their spot size as they propagate, contrary to the typical spreading out that characterizes wave phenomena.

This invariant behavior is described in terms of optical intensity by the relation:
\[
I(x, y, z > 0) = I(x, y, z = 0),
\]
where \( x \), \( y \), and \( z \) represent spatial coordinates. A notable example of this phenomenon is the "fundamental" Bessel beam, which features a transverse intensity distribution that is proportional to \( J_0^2(k_{\rho} \rho) \), where \( k_{\rho} \) is the wave number along the radial direction and \( \rho = \sqrt{x^2 + y^2} \) is the radial distance from the central axis of the beam.

\begin{figure}[t]
    \centering
    \includegraphics[width=\columnwidth]{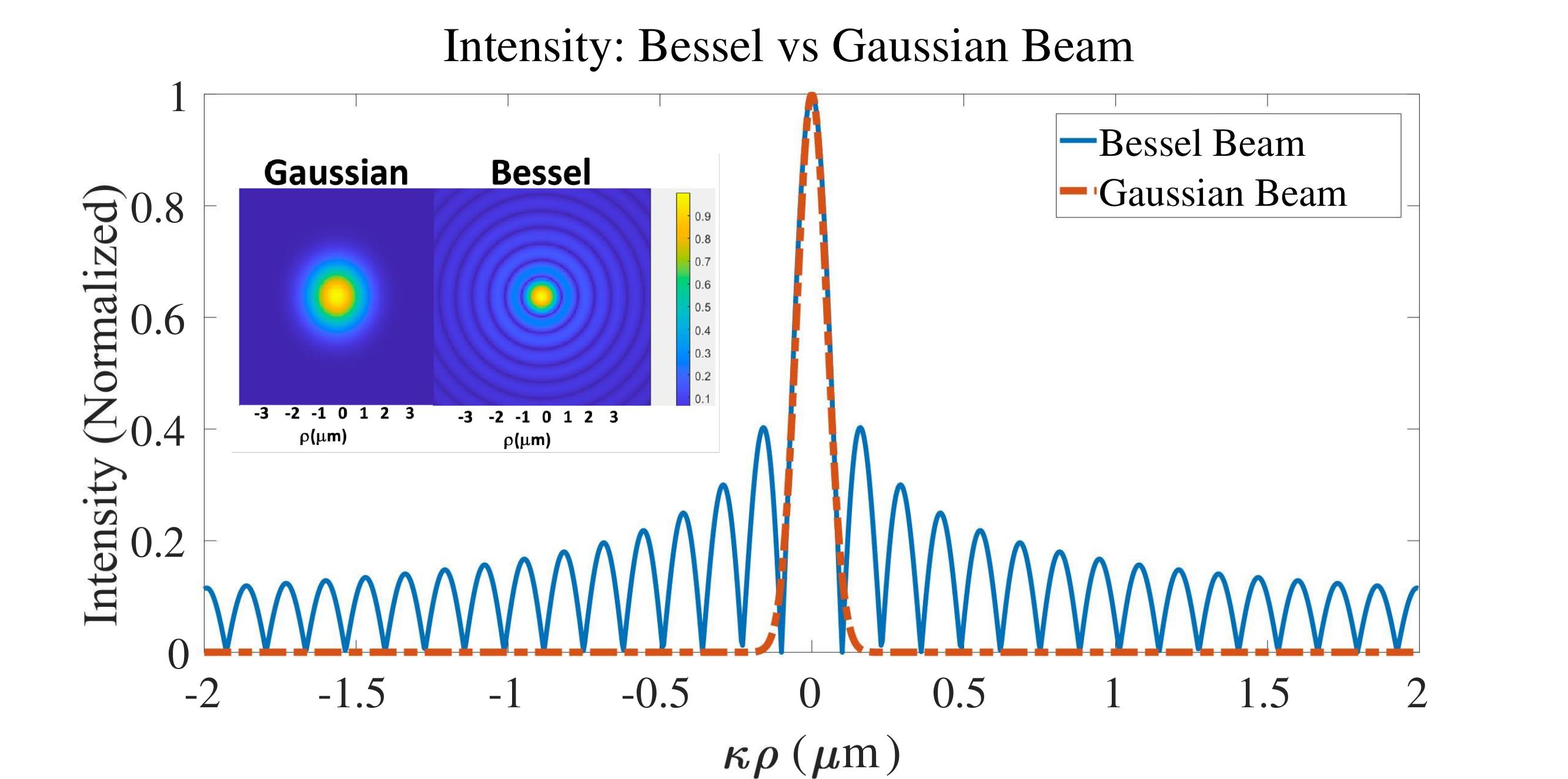}
    \caption{Normalized Intensity: Bessel Beam vs Gaussian Beam }
    \label{GsBs}
    
    % Define the coordinates for the inset
    \begin{tikzpicture}[remember picture,overlay]
        \coordinate (inset) at (-3,3); % Adjust the coordinates as needed
    \end{tikzpicture}
   
\end{figure}
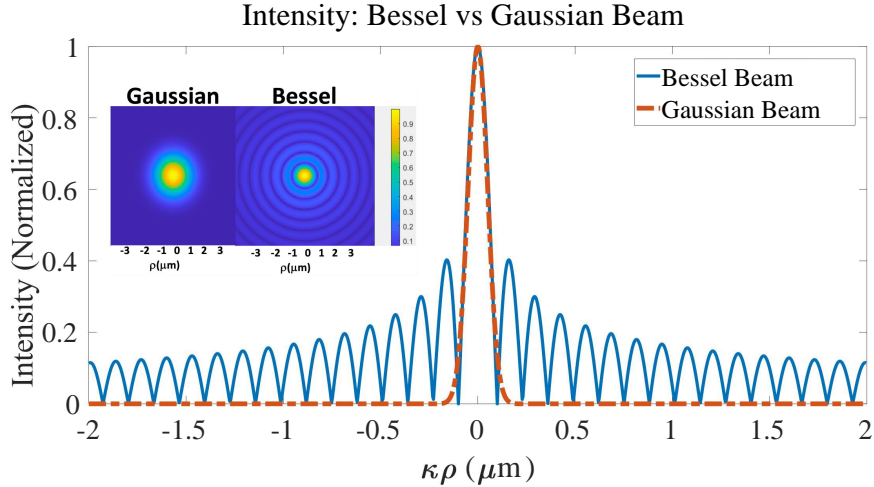

Although, in theory, Bessel beams are infinite in extent and energy, practical implementations using finite apertures can approximate these beams with a limited non-diffracting range (NDR) \cite{fuscaldo2023roadmap,pakovic2021bessel}. Compared to a Gaussian beam, the transverse intensity profile of Bessel beams remains consistent over distances that exceed the Rayleigh range of their Gaussian counterparts, making them prime candidates for applications requiring minimal wavefront distortion. Figure \ref{GsBs} shows the intensity comparison between the two beams. The extra side lobes in the Bessel beam mean that less energy is available in the main lobe. As such, the central lobe of the Gaussian beam contains more energy than that of a Bessel beam, albeit with a shorter non-diffracting range.

% Load the main image

The equation below shows the relationship between Z, the non-diffracting range, the transverse wavenumber $k_{\rho}$, and the radius R, of the emitting plane of the aperture. \cite{comite1999bessel,rao2024origin,khonina2020bessel,mcgloin2005bessel}:
 
\begin{equation}
Z_{\text{max}} = R \sqrt{\left[\left(\frac{k^2_0}{{k^2_\rho}} - 1\right)\right]} 
\label{eq:1}
\end{equation}

 where $Z_{max}$ is the maximum non-diffracting range and R is the finite radius of the emitting plane.

\subsection{Leaky waves in bounded structures: Spectral considerations}

\begin{figure}[b]
  \centering
  \includegraphics[width=\columnwidth]{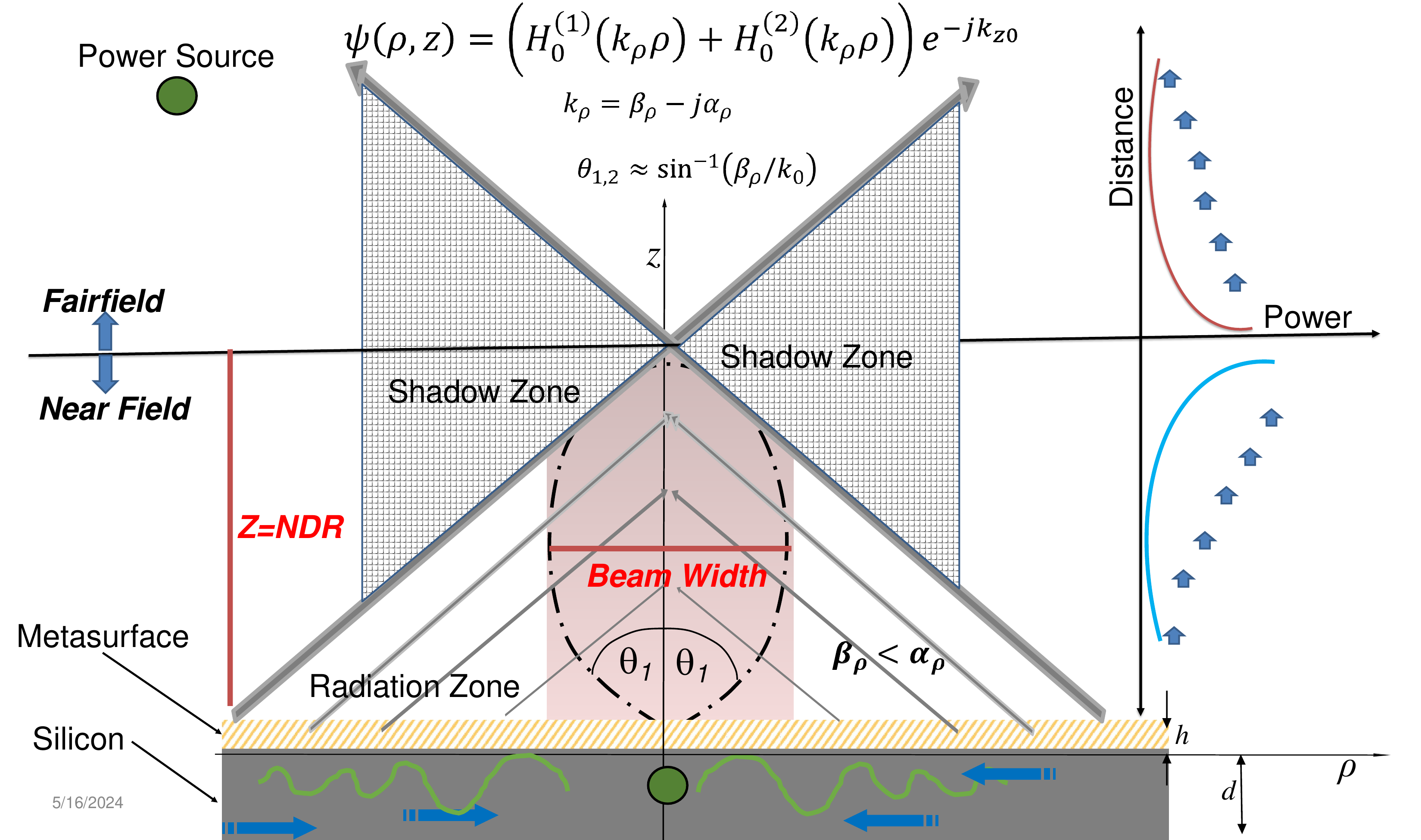} % Adjust the width to fit the column
  \caption{\label{fig:2} General Leaky Wave Structure with Radiation on a cone $\rho$ plane.}
\end{figure}

In the half-space above the emitting plane (see Fig. \ref{fig:2}), the wave number \(k_0\) is decomposed into transverse (\(k_{\rho}\)) and longitudinal (\(k_z\)) components. For radiation to occur, both \(k_z\) and \(k_{\rho}\) must be complex, with their real parts (\(\beta_{\rho}\) and \(\beta_z\)) defining the direction of propagation (\(\theta = \arcsin(\beta_{\rho} k_0^{-1})\)) and their imaginary parts representing attenuation constants. We note that \(k_x^2 + k_y^2 + k_z^2 = k_0^2\), where \(\boldsymbol{k_{\rho}} = \hat{x} k_x + \hat{y} k_y\) is the transverse wave number and \(k_z\) is the longitudinal wave number along the direction of propagation.
When \(k_{\rho} = 0\), the phase velocity equals the velocity of a plane wave in an unbounded medium, resulting in a TEM wave. Conversely, when \(k_{\rho} > 0\), the wave is a fast wave (leaky wave) along the \(z\) direction. The leaky wave component along \(z\) exists only in a defined portion of space (NDR) and does not exist in the shadow zone. The real part of the wave number, \(\beta_{\rho}\), determines the direction of propagation (forward or backward), while the imaginary part, \(\alpha_{\rho}\), determines the efficiency and power flow out of the structure.

A grating coupler (GC) is a typical leaky wave structure used to couple light in and out of a photonic chip. The GC typically uses a forward leaky wave, whereas a Bessel beam would require two backward traveling Hankel waves in the radiation zone as shown in Fig.~\ref{fig:2}. The complex wave numbers $k_\rho$ and $k_z$ can be expressed as:

\begin{subequations}
\label{eq:k0}
\begin{align}
    k_z &= \beta_z - j\alpha_z= k_0\sin{\theta_{air}}\label{eq:kz} \\
    k_{\rho} &= \beta_{\rho} - j\alpha_{\rho}= k_0\cos{\theta_{air}} \label{eq:krho} \\
    k_0 &= \pm\sqrt{{k^2_z} + {k^2_{\rho}}} \label{eq:2c}
\end{align}
\end{subequations}

Eq.~(\ref{eq:2c}) is multi-valued and the sign is chosen to satisfy Sommerfeld's radiation condition \cite{schot1992eighty}. For a leaky wave propagating forward, the value of $\beta_{\rho}$ is real and positive. To achieve a backward leaky wave with negative $\beta_{\rho}$, the structure's homogeneity is broken by introducing periodic scattering elements on top of the silicon slab. The phase constant $\beta_{\rho}$ of a periodic structure consists of space harmonics in accordance with Floquet expansion theory \cite{keqian2001electromagnetic}:
 
 \begin{equation}
 \label{eq:betan}
     \beta_{\rho,n}=\beta_{\rho}+\frac{2\pi{n}}{a}=k_0\sin{\theta_n}
 \end{equation}
 
where a is the period of the unit cell. The negative spatial harmonics with $n=-1,-2,...$ radiate backward with a suitable choice of $a$. 

\section{Derivation of Electric Fields in Bounded Structures}

\begin{figure} [b] \label{cartesian}
 \centering
  \includegraphics[width=\columnwidth]{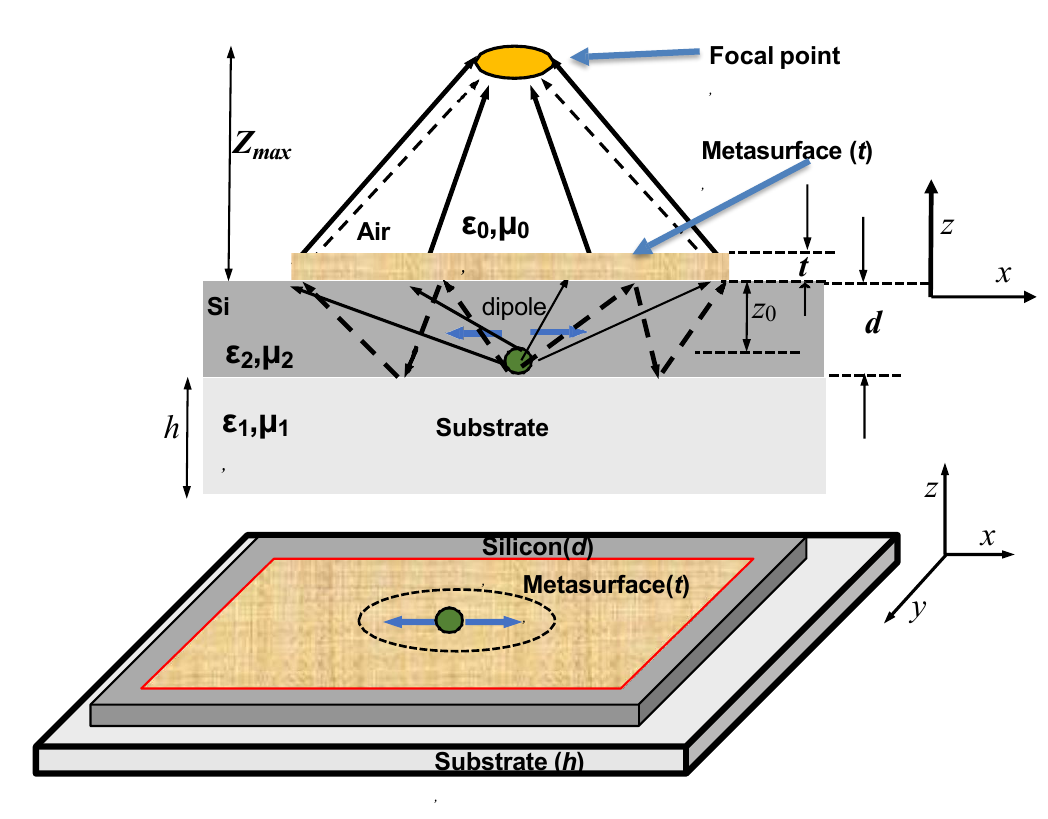} % Adjust the width to fit the column
\caption{\label{fig:3} Schematic of the dielectric stack with the light source placed in the silicon region.}
\end{figure}

In this section, we derive the electric fields in the various regions of the structure in Fig. \ref{fig:3} by use of Dyadic Green's functions. We assume an EM source as an electric dipole. Green's function describes the system's impulse response.

To solve the inhomogeneous differential equation with a source term, typical methods include the use of the Wronskian determinant 
\cite{jackson2012classical,park1995fokker,economou2006green,schwartz2012principles}, the spectral domain method, also referred to as the Sturm-Liouville approach \cite{dudley1994mathematical,collin1990field,angell1985helmholtz,felsen1994radiation} and the method of images for symmetric boundaries \cite{tai1971dyadic,duffy2015green,stakgold2011green}. Additional methods include integral transforms such as Fourier and Laplace \cite{collin1990field}, variational techniques \cite{chew1999waves}, perturbation approaches like WKB for non-linear operators \cite{chew1999waves,ishimaru2017electromagnetic}, finite element analysis for complex geometries, and stochastic methods like Monte Carlo simulations for statistical estimation \cite{tsang2004scattering,wagner1997monte}.

We particularly focus on the spectral method for its usefulness when dealing with self-adjoint operators. The following subsections detail our approach:

\begin{enumerate}[label=\alph*)]
    \item Begin with a single vector potential function to generate orthogonal eigenfunctions for TE and TM waves, represented by vectors $\mathbf{M}$, $\mathbf{N}$, and $\mathbf{L}$.
    \item Solve for the free space magnetic (${G}^m$) and electric (${G}^e$) dyadic Green's functions in homogeneous media with a dipole source.
    \item Implement boundary conditions and apply the superposition principle to address Green's functions in the half-space and substrate regions.
    \item Utilize the steepest descent contour (SDC) technique for the inverse Fourier transform of spectral fields back into the spatial domain.
\end{enumerate}

We start by defining a unified potential function from which a complete set of orthogonal eigenfunctions will be derived \cite{tai1971dyadic}.

\subsection{Vector wave functions}
The scalar wave function $\psi$ in the homogeneous silicon region is written as follows:

\begin{equation}
\label{eq:4a}
\psi = \hat{z} e^{-j\mathbf{k}\cdot\mathbf{R}} = \hat{z} e^{-j(k_x x + k_y y + k_z z)}
\end{equation}

Using $\psi$, as the generating function, we derive a set of orthogonal vector wave functions $\mathbf{M}$, $\mathbf{N}$ and $\mathbf{L}$  that satisfy the vector wave equation.

\cite{collin1990field,chew1999waves,tai1971dyadic}:
\begin{equation}
    {\nabla}\times{\nabla}\times{\mathbf{F}}+{k^2}\mathbf{F}=0
\end{equation}
 The basis vector wave functions $\mathbf{F}( \mathbf{M}, \mathbf{N})$ are derived by the following curl ($\nabla \times$) operations: 
 
\begin{subequations}
\label{eq:whole:8}
\begin{align}
\mathbf{M} &= \nabla \times (\psi) 
\label{subeq:8a} \\
\mathbf{M'} &= \nabla \times (\psi) 
\label{subeq:8b} \\
\mathbf{N'} &= \nabla \times \nabla \times (\psi) 
\label{subeq:8c} \\
\mathbf{N} &= \nabla \times \nabla \times (\psi) 
\label{subeq:8d} \\
\mathbf{L} &= \nabla (\psi) 
\label{subeq:8e}
\end{align}
\end{subequations}

The primed functions $\mathbf{M'}$ and $\mathbf{N'}$ are fixed vector functions defined with respect to the source point. The wave functions also satisfy the orthogonality relation below:

\begin{align}
\iiint_v \mathbf{M}(k) \cdot \mathbf{M'}(-k) dV = \iiint_v \mathbf{N'}(k) \cdot \mathbf{N}(-k) dV  
= (2\pi)^3 (k_x^2 + k_y^2) \delta (k - k') \label{eq:11}
\end{align}

\subsection{$G^m$ and $G^e$ free space dyadic Green's Functions}

The free-space dyadic Green's functions (DGFs) in the silicon region satisfy the following differential equations and will later be expressed in terms of the eigenfunctions $\mathbf{M}$ and $\mathbf{N}$.

\begin{subequations} \label{Gall1}
  \begin{align}
    \nabla \times \nabla \times {G}^e_{si}(\mathbf{R}, \mathbf{R}') - k^2 {G}^e_{si}(\mathbf{R}, \mathbf{R}') &= \overline{\overline{I}} \delta(\mathbf{R} - \mathbf{R}') \label{Gall1a} \\
    \nabla \times \nabla \times {G}^m_{si}(\mathbf{R}, \mathbf{R}') - k^2 {G}^m_{si}(\mathbf{R}, \mathbf{R}') &= \nabla \times \overline{\overline{I}} \delta(\mathbf{R} - \mathbf{R}') \label{Gall1b}
  \end{align}
\end{subequations}

The delta function $\overline{\overline{I}}\delta(\overline{R}-{\overline{R'}})$ on the right-hand side represents the forcing function of unit magnitude and is located at an arbitrary position vector $\boldsymbol{{{R'}}}$. The vector $\boldsymbol{{{R}}}$ is the position of the observer, and $\overline{\overline{I}}$ is the idem factor. 
The GFs in the other two regions, i.e. free space and substrate regions, are written without the source term as:
\begin{subequations}
\label{Gall2}
  \begin{align}
    \nabla\times\nabla\times {G}^e_{sub}(\mathbf{R}, \mathbf{R}') - k^2 {G}^e_{sub}(\mathbf{R}, \mathbf{R}') &= 0 
    \label{12c} \\
    \nabla \times\nabla \times {G}^e_{air}(\mathbf{R}, \mathbf{R}') - k^2 {G}^e_{air}(\mathbf{R}, \mathbf{R}') &= 0
    \label{12d}
  \end{align}
\end{subequations} \\
We solve for ${G^m_{si}}$ in Eq.~(\ref{Gall1b}) first and then use the relation $\nabla\times{{G}^m}={G}^e$ to solve for ${G}^e$. From the basic foundations of Maxwell's equations, a complete definition of the electric fields would need solenoidal ($\nabla\times$) and irrotational ($(\nabla\cdot)$) components, whereas the magnetic fields are divergence-free ($\nabla\cdot\boldsymbol{B=0}$) and require only solenoidal components. Therefore, ${G}^m$ requires just two eigenfunctions ($\mathbf{M}$ and $\mathbf{N}$).

Equation ~(\ref{Gall1b}) is then expanded in terms of orthogonal basis functions $\mathbf{M}$ and $\mathbf{N}$ with scaling coefficients A,B,C and D to be determined. The right-hand side is expanded as
 
\begin{equation} 
\label{delAB}
\nabla \times \overline{\overline{I}} \delta(\mathbf{R} - \mathbf{R'}) = \iiint_{-\infty}^{\infty} [\mathbf{M}(k) A(k) + \mathbf{N}(k) B(k)] \, d\kappa
\end{equation}

where $d\kappa=dk_x \, dk_y \, dk_z$

$G^{m}$ on the left-hand side is expanded as:

\begin{align} \label{delCD}
G^{m}(\mathbf{R}, \mathbf{R}') = \frac{1}{(2\pi)^3}\iiint_{-\infty}^{\infty} dk_x dk_y dk_z  \left[ C\mathbf{N}(k) + D\mathbf{M}(k) \right]
\end{align}

${A}(k)$ and ${B}(k)$ are solved by multiplying both sides of the equation by the  functions $\mathbf{M'}({k})$ and $\mathbf{N'}({k})$ respectively. Eq. (\ref{delAB}) now becomes:

\begin{equation}
\begin{aligned}
\nabla\times\overline{\overline{I}}\delta(\overline{R}-{\overline{R'}}) = \iiint_{-\infty}^{\infty} \frac{k({k_x}^2+{k_y}^2)}{2(\pi)^3} 
\times [\mathbf{N}({k})\mathbf{M'}({-k})+\mathbf{M}({k})\mathbf{N'}({-k})] \,d\kappa
,¬\end{aligned}  
\label{eq:whole:12_delta}
\end{equation} \\
After solving for $C(k)$ and $D(k)$, Eq. (\ref{delCD}) is integrated with respect to $k_z$ using the Cauchy residual theorem (CRT) which reduces to (Algebra omitted for brevity):
\begin{align} \label{delcom1}
 G^{m}(\mathbf{R}, \mathbf{R}') = \frac{-j}{(2\pi)^2}\iint_{-\infty}^{\infty} dk_xk_y \frac{1}{(k_x^2+k_y^2)k_z'} 
 \times \left[ \mathbf{N}(k) \mathbf{N'}(\pm k_z') + \mathbf{M}(k) \mathbf{M'}(\pm k_z') \right]
\end{align}

Eq.~(\ref{delcom1}) is the free space magnetic dyadic GF a discontinuity in the $k_z$ plane. To obtain $G^e$ from $G^m$, we use the relation $\nabla\times{G^m}=G^e$, and put $G^e$ in evidence from Eq.~(\ref{Gall1a}) to yield:

\begin{equation} \label{eq:G_el}
G^{e}(\mathbf{R}, \mathbf{R}') = \frac{1}{k^2} \left[- \overline{\overline{I}} \delta(\mathbf{R} - \mathbf{R}') + \nabla \times G^{m}(\mathbf{R}, \mathbf{R}') \right]
\end{equation}

Further simplification of $G^e$ in Eq. (\ref{eq:G_el}) gives:

\begin{align} \label{eq:G_el2}
G^{e}(\mathbf{R}, \mathbf{R}') &= 
  \frac{1}{k^2} \left[ -\hat{z} \hat{z} \delta(\mathbf{R} - \mathbf{R}') + \right. \notag \\ 
 & \left. \frac{-j}{(2\pi)^2} \iint_{-\infty}^{\infty} dk_x dk_y \frac{1}{(k_x^2+k_y^2)k_z'} 
 \times \left( \mathbf{N}(\pm k_z') \mathbf{N'}(\mp k_z') + \mathbf{M}(\pm k_z') \mathbf{M'}(\mp k_z') \right) \right]
\end{align}

The divergence-free nature of the magnetic field $(\nabla\cdot\mathbf{B}=0)$ suppresses the delta function in ${G^m}$ but it appears in the ${G^m}$ to account for the nonsolenoidal ($\mathbf{L}$) electric fields. It has been customary to derive the $G^e$ functions directly using Hertz vector potentials and mixed potential functions without the need for ${G}^m$ \cite{collin1990field,stratton2007electromagnetic, chew1999waves,michalski1990electromagnetic}. However, as discussed in \cite{tai1971dyadic}, they all require three vector functions instead of two, which makes the computation arduous.

\subsection{Dyadic Green's Function $G^e$ in bounded media} \label{Dyad}

We now return to the multilayered structure in Fig.~\ref{fig:3} and use the principle of superposition to define the respective GFs in the substrate and air regions. For clarity we redefine ${G^e}$ to mean the modified GF to be derived in the bounded medium and ${G}^e_{0,si}$ to be the primary GF in a homogeneous medium derived in Eq. (\ref{eq:G_el2}):

\begin{align} \label{eq:G_total}
& G^{e} = \begin{cases} 
0  \\
G_{0,si}^e(\mathbf{R}, \mathbf{R}')  \\
0 
\end{cases}+ \\ \notag
& \iint_{0}^{\infty}dk_xk_y\frac{-j}{4\pi{(k_x^2+k_y^2)k_z'}}
\begin{cases} 
e\mathbf{M}(k_1)+f\mathbf{N}(k_1), & z \geq 0 \\
a\mathbf{M}(k_2) +b\mathbf{M}(-k_2) +c\mathbf{N}(k_2) +d\mathbf{N}(-k_2) , & 0 > z > -d \\
g\mathbf{M}(-k_3)+h\mathbf{N}(-k_3), & z < -d
\end{cases}
\end{align}    

The constants $a,b,c,d,e,f,g,h$ are the amplitude coefficients at the media interfaces. They are obtained by imposing the continuity of both electric and magnetic components across boundaries and are listed in Appendix A. The GF in the air region is written below whereas the GFs for substrate and silicon region are  listed in Appendix B 

\begin{equation}
\begin{split}
G^{e}_{air}(\mathbf{R},\mathbf{R'}) = & \iint_{0}^{\infty} dk_x dk_y \left(\frac{-j}{4\pi(k_x^2 + k_y^2)k_z'}\right) \bigg\{ \mathbf{M}(k_{1z}) \frac{T^{TE}_1}{W^{TE}} \left[ R^{TE}_2 e^{-j2\phi_2} \mathbf{M'}(k_{2z}) + \mathbf{M'}(-k_{2z}) \right] \\
& + \mathbf{N}(k_{2z}) \frac{T^{TM}_1}{W^{TM}} \frac{k_{si}}{k_{air}}\left[ R^{TM}_2 e^{-j2\phi_2} \mathbf{N'}(k_{1z}) + \mathbf{N'}(-k_{1z}) \right] \bigg\}
\end{split}
\label{eq:GreenFunctions}
\end{equation}

$W^{TE}$ and W$^{TM}$ in the denominator define the phase matching condition at the boundaries for TE and TM fields (Defined in Appendix A). $T^{TE}_1$ is the transmission coefficient between silicon and air whereas $R^{TE}_2$ defines the reflection coefficient at the substrate-silicon interface. The double integral in Eq. \ref{eq:GreenFunctions} can be solved numerically. However, we derive its closed-form solutions to gain insight into the leaky wave phenomenon. This is done by recasting the double integrals into Sommerfeld type integral in cylindrical coordinates $({\rho,\phi,z})$. 
To proceed, we also adopt the Fourier-Bessel decomposition of the free space DGF \cite{chew1999waves,stratton2007electromagnetic,ishimaru2017electromagnetic} in which the $\phi$ plane (azimuthal plane) is written as a Fourier series while the $\rho$ plane is written as a Fourier transform. We now have:

\begin{equation}
\begin{aligned}
G_0(\mathbf{R, R'}) = \frac{1}{4\pi} \sum_{m=-\infty}^{\infty} e^{-jm(\phi - \phi')} 
 \times \int_{0}^{\infty} J_m(k{\rho})J_m(k{\rho'}) e^{-jk_z|z-z'|} \frac{k_\rho}{jk_z} d\rho \\
\end{aligned}
\label{eq:Cy}
\end{equation}

And Eq. \ref{eq:GreenFunctions} for $\mathbf{G^e_{air}}$ now becomes:

\begin{equation}
\begin{aligned}
B &= \frac{-j(2-\delta)}{4\pi k_{\rho} k_z'}, \\
k_{\rho}^2 &= k_0^2 - k_{z1}^2, \\
G^{e}_{\text{air}}(\mathbf{R},\mathbf{R'}) &= \int_{0}^{\infty} d\rho \sum_{m=0}^{\infty} B \Bigg\{ \mathbf{M}(k_{1z}) \frac{T^{TE}_1}{W^{TE}} \left[ R^{TE}_2 e^{-j2\phi_2} \mathbf{M'}(k_{1z}) + \mathbf{M'}(-k_{2z}) \right] \\
&\quad + \mathbf{N}(k_{2z}) \frac{T^{TM}_1}{W^{TM}} \left( \frac{k_{\text{si}}}{k_{\text{air}}} \right) \left[ R^{TM}_2 e^{-j2\phi_2} \mathbf{N'}(k_{1z}) + \mathbf{N'}(-k_{1z}) \right] \Bigg\}.
\end{aligned}
\label{GF_cy}
\end{equation}

In Eq.(\ref{GF_cy}), $\delta=0$ for $n>0$ and the Cartesian eigen vectors $\mathbf{M}$ and $\mathbf{N}$ are transformed into their equivalent cylindrical vector functions. Thus:
\begin{subequations}
\begin{align}
\mathbf{M}_{eon}(k_{z1}) &= \nabla \times [\psi_{eonk_{\rho}}(k_{z1}) \hat{z}] \\
\mathbf{M}_{eon}(k_{z1}) &=
\begin{bmatrix}
\pm\frac{nJ_n(k_{\rho} \rho)}{\rho} \begin{cases}
    \sin(n\phi) \\
    \cos(n\phi)
\end{cases} \\
\frac{J_n(k_{\rho} \rho)}{\rho} \begin{cases}
    \cos(n\phi) \\
    \sin(n\phi)
\end{cases} \\
0
\end{bmatrix} e^{-jk_{z1}z} \\
\mathbf{N}_{eon}(k_{z1}) &= \frac{1}{k_{1}} \nabla \times \mathbf{M}_{eon}(k_{z1}) \\
\mathbf{N}_{eon}(k_{z1}) &= \frac{1}{k_{1}}
\begin{bmatrix}
-jk_{z1} \frac{\partial J_n(k_{\rho} \rho)}{\partial \rho}\begin{cases}
    \cos(n\phi) \\
    \sin(n\phi)
\end{cases}  \\
\pm j \frac{nJ_n(k_{\rho} \rho)}{k_{\rho} \rho} \begin{cases}
    \sin(n\phi) \\
    \cos(n\phi)
\end{cases} \\
k^2_{\rho} J_n(k_{\rho} \rho) \begin{cases}
    \cos(n\phi) \\
    \sin(n\phi)
\end{cases}
\end{bmatrix} e^{-jk_{z1}z}
\end{align} 
\end{subequations}

Where the subscript symbols ${eon}$ stand for {\textbf{even}, \textbf{odd}} functions of cosine and sine, respectively, and $n$ for the harmonic number. The subscripts ${eon}$ are also implied in Eq. (\ref{GF_cy}) but suppressed for clarity.
The other components of the DGF in the silicon (Si) and substrate (Sub) regions can be written in a similar manner but have been omitted here for brevity.

\subsubsection{\textbf{Extracting Electric Field from Greens Functions}}

The electric field in the different regions can now be written using the derived GFs as:

\begin{subequations}
\begin{align}
\mathbf{E}_{\text{air}}(\mathbf{R}) &= \iiint_{-\infty}^{\infty} \mathbf{G}_{\text{air}}(\mathbf{R},\mathbf{R'}) \cdot \mathbf{J}_{\text{si}}(\mathbf{R'}) \, dV', \label{eq:air} \\
\mathbf{E}_{\text{si}}(\mathbf{R}) &= \iiint_{-\infty}^{\infty} \mathbf{G}_{\text{si}}(\mathbf{R},\mathbf{R'}) \cdot \mathbf{J}_{\text{si}}(\mathbf{R'}) \, dV', \label{eq:si} \\
\mathbf{E}_{\text{sub}}(\mathbf{R}) &= \iiint_{-\infty}^{\infty} \mathbf{G}_{\text{sub}}(\mathbf{R},\mathbf{R'}) \cdot \mathbf{J}_{\text{si}}(\mathbf{R'}) \, dV'. \label{eq:sub}
\end{align} \label{Ef}
\end{subequations}

The current source in Eq. \ref{Ef}, is an electric dipole located at $\mathbf{R'}$ with a dipole moment oriented in one of the Cartesian planes $\hat{x},\hat{y},\hat{z}$. The dipole moment is chosen as $4\frac{j4{\pi}k^2_{\rho}}{\omega\mu_0}$ to normalize the constant B in Eq. \ref{GF_cy}. Thus,

\begin{equation}
\mathbf{J}_{\text{si}}(\mathbf{R'})=\frac{j4{\pi}k^2_{\rho}}{\omega\mu_0}(\hat{x},\hat{y},\hat{z})\delta(x'-x)\delta(y'-y)\delta(z'-z)    
\end{equation}

We will consider a linearly polarized dipole at $R=(0,0,z_0)$, oriented along the $x$ direction at the center of the silicon slab waveguide. According to Eq. \ref{Ef}, the electric field in the air region above the waveguide becomes: 
\begin{equation}
    \mathbf{E}_{\text{air}} =  \mathbf{G}_{\text{air}}(\mathbf{R},\mathbf{R'}) \cdot {\hat{x}} \label{Ef2} \\
\end{equation}
We note that $\hat{x}$ is 
\begin{equation*}
   \hat{x}=\hat{\rho}\cos{\theta}-\hat{\phi}\sin{\theta} 
\end{equation*}
 The components of the vector wave functions that emerge from the dot product operation are $M_{01k_{\rho}}(k_{z1})$ and  $N_{e1k_{\rho}}(k_{z1})$. Therefore

\begin{flalign}
& E_{\text{air}}(\mathbf{R})= \frac{T^{TE}_1}{W^{TE}} 2k_1 \\
&  \times \int_{0}^{\infty} \Bigg[ \frac{k_{2z}}{k_{1z}} \mathbf{M}_{01 k_\rho}(k_{1z}) \left( e^{jk_{2z}z_0} + R^{TE}_1 e^{-jk_{2z}z_0} \right)  
+ j \frac{T^{TM}_1}{W^{TM}} \left(\frac{k_{\text{si}}}{k_{\text{air}}}\right) \mathbf{N}_{e1 k_\rho}(k_{1z}) \left( e^{-jk_{2z}z_0} \right) \Bigg] \, dk_{\rho} && \notag \\
&\quad - j \frac{T^{TM}_1}{W^{TM}} \left(\frac{k_{\text{si}}}{k_{\text{air}}}\right) \mathcal{N}_{e1 k_\rho}(k_{1z}) R^{TM}_2 e^{-j2\phi_2} e^{-jk_{2z}z_0} \, dk_{\rho} &&
\label{E_air}
\end{flalign}

The following expansions have been used in the computation:

\begin{subequations}
\begin{align}
\frac{d[J_n(k_{\rho}\rho)]}{d\rho} &= \frac{1}{2}\left[J_{n+1}(k_{\rho}\rho) + J_{n-1}(k_{\rho}\rho)\right] \\
[J_n(k_{\rho}\rho')] &= 0; \quad n > 0, \rho' = 0 \\
[J_0(k_{\rho}\rho')] &= 1; \quad n = 0
\end{align}
\end{subequations}

\subsection{Solving for the Electric Fields in Half Space using the SDC technique}

In order to Evaluate the integral in Eq. (\ref{E_air}), we use an asymptotic method that combines Laplace's method of integrals and the method of stationary phase as detailed in Bender and Orszag (pg. 261-305) \cite{bender2013advanced}. This method is used to approximate integrals of the form:
\[
I = \int_{-\infty}^{\infty} f(x) e^{-j \lambda g(x)} \, dx
\]
where \( f(x) \) and \( g(x) \) are real-valued functions, and \( \lambda \) is a large positive parameter. The goal is to evaluate the integral for large values of \( \lambda \) by identifying the dominant contribution to the integral.
The critical points of the phase function \( g(x) \) offer the dominant contribution to the integral, since at this point the phase is approximately constant. 
Following this procedure, the integral in Eq. \ref{E_air} is reshaped into the form of $I$. A key difference is that the function $f(x)$ has pole singularities to be considered. The integral has a continuous spectrum away from the poles and has a discrete spectrum around the poles. The contribution of the discrete spectrum is given by equating the denominator $W^{TE,TM}$ of the GF to zero. This would yield contributions from leaky wave poles and surface wave poles \cite{ishimaru2017electromagnetic}. The continuous spectrum is obtained by integrating around a suitable saddle point by making the following transformations.

\begin{subequations}
\begin{align}
z - d &= R \sin \theta, \quad \rho = R \cos \theta, \label{eq:zdR}\\
k_{\rho} &= k_0 \sin \Phi, \label{eq:krho}\\
k_z &= k_0 \cos \Phi, \label{eq:kz}\\
% You should define K(\alpha) here if it's used later
E_{\text{air}} &= \int_C K(\Phi) \exp[k_0 r f(\Phi)] d\Phi, \label{eq:G_air} \\
K(\Phi) &= \frac{A (\Phi) k_0 \cos \Phi}{2\pi}, \label{eq:KPhi} \\
f(\Phi) &= -j \cos(\Phi - \theta). \label{eq:fPhi}
\end{align}
\end{subequations}

Noting that the integral is a sum of four sub-integrals. We evaluate the first two containing $\mathbf{M}$ functions.

\begin{subequations}\label{I12}
\begin{equation}
I_1 = \frac{1}{4\pi} \int_0^{\infty} \left[ \frac{T^{TE}_1 e^{-\left(jk_{z2}(d-z_0) + k_{z1}(z-d)\right)}}{1 - R^{TE}_1 R^{TE}_3 e^{-2jk_{z2}d}} \right] \times J_0(k_{\rho} \rho) k_{\rho} \frac{dk_{\rho}}{jk_{z2}} \label{I12a}
\end{equation}
\vspace{2ex} % Adjust the space as needed
\begin{equation}
I_2 = \frac{1}{4\pi} \int_0^{\infty} \left[ \frac{T^{TE}_1 R^{TE}_3 e^{-\left(jk_{z2}(d+z_0) + k_{z1}(z-d)\right)}}{1 - R^{TE}_1 R^{TE}_3 e^{-2jk_{z2}d}} \right] \times J_0(k_{\rho} \rho) k_{\rho} \frac{dk_{\rho}}{jk_{z2}} \label{I12b}
\end{equation}
\end{subequations}

The quantities in brackets with the integral of Eq. \ref{I12} can be written:

\begin{equation}
S1 \equiv \sum_{n=0}^{\infty} U_{n}, \quad S2 \equiv \sum_{n=0}^{\infty} V_{n}
\label{eq:S_definitions}
\end{equation}

Where
\begin{subequations} \label{UV}
\begin{equation}
U_n = R^{TE}_1 e^{-j k_{z2}(d-z_0)} e^{-j k_{z1} (z-d)} 
\times (R^{TE}_1 R^{TE}_3)^n e^{-2jnk_{z2}d} \label{eq:Un}  \end{equation}
%\vspace{1ex}    
\begin{equation}
V_n = T^{TE}_1 R^{TE}_1 e^{-j k_{z2}(d-z_0)}
 \times e^{-j k_{z1} (z-d)} (R^{TE}_1 R^{TE}_3)^n e^{-2jnk_{z2}d} \label{eq:Vn}
\end{equation}
   \end{subequations}

Each of the wave components in Eq. \ref{I12} can be expressed in the form of the saddle point technique in  Eq. \ref{eq:fPhi} and evaluated using the saddle point method as below:

\begin{multline}
\int_{0}^{\infty} A(\rho)e^{-j k_{z2}(d-z_0)} J_0 (k_{\rho} \rho) \rho d\rho \\ \approx
\frac{1}{8\pi} \int_{-\infty}^{\infty} A(\rho)e^{-j k_{z2}(d-z_0)}H_0^{(2)}(k_{\rho} \rho) \frac{\rho dk_{\rho}}{j k_{z1}} \\ \approx 
\frac{1}{8\pi} \int_{-\infty}^{\infty} \left[ \sqrt{\frac{2}{\pi k_{\rho} \rho}} A(k_{\rho}) \frac{k_{\rho} \rho e^{j\pi/4}}{j k_{z2}} \right]e^{-jf(k_{\rho})} dk_{\rho}
\label{SPT1}
\end{multline}

In Eq. \ref{SPT1}, 
\begin{equation}
f(k_{\rho}) = k_{z2}d + k_{z1}z + k_{\rho} \rho
\label{eq:f_function}
\end{equation}

Transforming the wave numbers $k_{\rho},k_{z1}$ from Cartesian coordinates to cylindrical coordinates and then integrating yields

\begin{equation}
\left[ \sqrt{\frac{2}{\pi k_{\rho} \rho}} A(k_{\rho}) \frac{k_{\rho} \rho e^{j\pi/4}}{j k_{z2}} \right] \sqrt{\frac{2\pi}{-f''(k_{\rho})}} 
\label{eq:last_on_page}
\end{equation}

\begin{figure}[t]
\centering

\hspace*{1em} \includegraphics[width=0.7\textwidth]{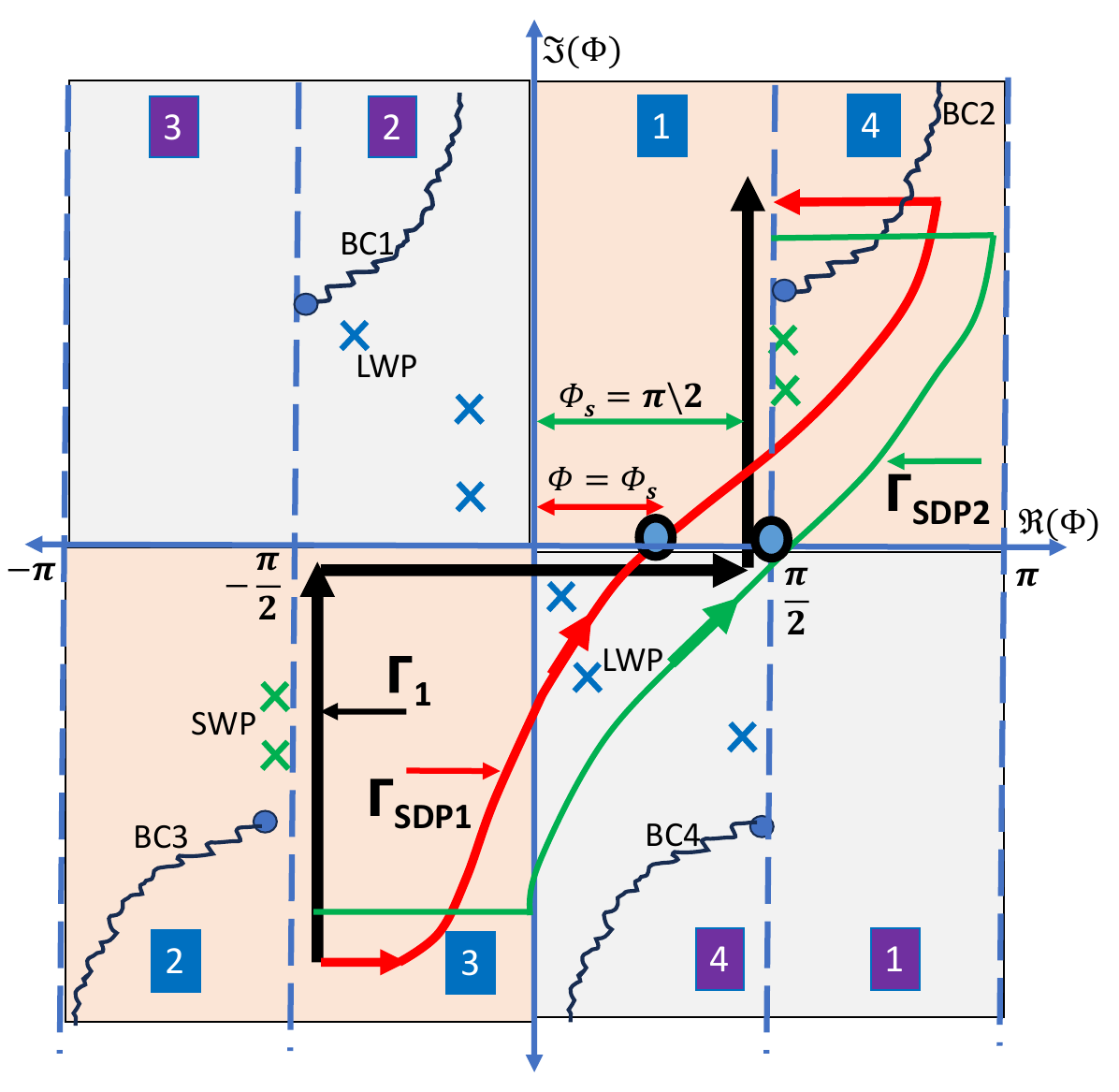}

\caption{Contour Integration mapping of the the complex plane $\Phi$ plane for evaluating the electric field.}
\label{fig:contour_around_branch_cuts2}
\end{figure}

The analysis of Eq.~\ref{eq:last_on_page} is performed using the steepest descent contour plot, as depicted in Fig.~\ref{fig:contour_around_branch_cuts2}, which provides a robust framework for evaluating complex integrals involving multi-valued functions via Sommerfeld radiation integrals. This plot highlights the deformation of the original integration path, marked by a dark solid line along $\pm \pi/2$, to optimize the integral evaluation paths through saddle points \(\Phi\), which denote regions of minimal phase change. 

Distinct colors indicate the transition between two Riemann sheets—brown for the upper sheet and gray for the lower sheet—representing different branches of the $\pm k_{z1}$ function, continuous within each sheet but varying between them. The contour paths, \(\Gamma_{(1,2)}\), are specifically designed to pass through these saddle points, depending on the location of the leaky wave pole (LWP) and surface wave pole (SWP) contributions. The red and green lines denote two viable paths for the integral evaluation, where the red line is aligned with the saddle point for leaky wave angles \(\Phi_s\), and the green line corresponds to the surface wave paths along the structure.
Branch cuts labeled as BC1, BC2, BC3, and BC4 are strategically placed to avoid discontinuities during the integration process.

\section{Introducing the Metasurface on the reference structure} \label{seC}

 \begin{figure}[b]
  \centering
 \hspace*{-1em} \includegraphics[width=0.8\textwidth]{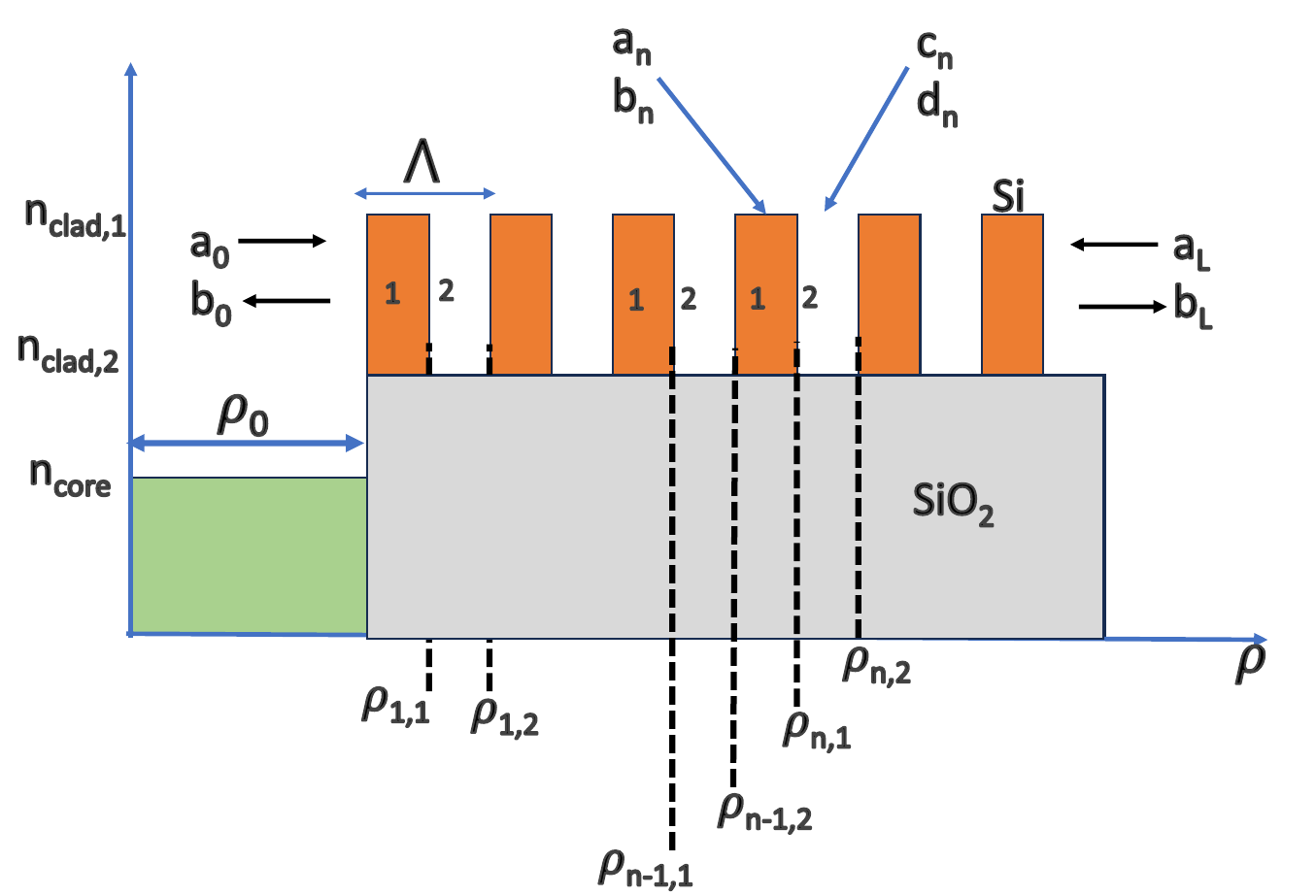} % Adjust the width to 1.2 times the text width
  \caption{Schematic of the metasurface cross section on top of a substrate ($n_{\text{clad}}$) with alternating cladding layers of thicknesses $t_1$ and $t_2$. The core radius is $\rho_0$ and $\Lambda$ is the period.}
  \label{fig:bragg}
\end{figure}

In this section, we introduce a perturbation on the three-layer reference surface in form of a periodic metasurface.  We study the periodic structure using the Bloch theorem, which is not commonly applied to cylindrical periodic structures. The structure is shown in Fig. 5 with a low index core ($n_l$) and pairs of low refractive index cladding ($n_1$ and $n_2$). 
The material in the perturbed medium has periodicity \( \Lambda \), and the propagating wave is represented by \:

\begin{equation} \label{blo}
    \sqrt{\rho+\Lambda}F(\rho + \Lambda) = \exp(-jK\Lambda)\sqrt{\rho}F(\rho)
\end{equation}

Eq. (\ref{blo}) is obtained by neglecting the first-order term \(\frac{1}{\rho}\frac{\partial F(\rho)}{\partial \rho}\) in the Bessel differential equation \cite{collin1990field}.  
\( K \) is the complex Bloch number for the periodic lattice and \( \Lambda \) is the period. \( F(\rho) \) is generally periodic and represents any of the fields \( E_{\theta}, H_{\theta}, E_{\rho}, H_{\rho} \) in the medium. Considering TE as an example, the starting potential $``U"$=$E_z$ field is $a_0J_m(k\rho)(\cos m\theta, \sin m\theta)$. The fields in the different regions are:

\begin{subequations}\label{eq:main}
    \begin{align}
        U = E_z = a_0 J_m(k_{\rho}\rho)(\cos m\theta, \sin m\theta)\exp({-jk_z z}) 
       \quad \rho < \rho' \label{eq:U1}  \\
        E^{TE}_z = a_n H^{(1)}(k_{\rho_1}\rho) + b_n H^{(2)}(k_{\rho_1}\rho)  
        \quad \rho_n < \rho < \rho_n + t_1 \label{eq:U2} \\ 
        E^{TE}_z = c_n H^{(1)}(k_{\rho_2}\rho) + d_n H^{(2)}(k_{\rho_2}\rho) 
        \quad \rho_{n+1} < \rho < \rho_{n+1} + t_2 \notag \label{eq:U3} \\
    \end{align}
\end{subequations}

where $k_{\rho_0} = \sqrt{n_0^2 k_0^2 - k_z^2}$, $k_{\rho_1} = \sqrt{n_1^2 k_0^2 - k_z^2}$, and $k_{\rho_2} = \sqrt{n_2^2 k_0^2 - k_z^2}$. The Hankel function $H(k_{\rho}\rho)$ is used for traveling waves in cylindrical coordinates. 

From $U$($E^{TE}$), corresponding $H_z^{TE}$ and $E^{TE}_\theta$ fields are derived. Matching boundary conditions between adjacent layers ($n_{(clad,1)},n_{(clad,2)}$) in which $E_\theta$, $H_\theta$, $E_\rho$ and $H_\rho$ must be continuous, we express the amplitude coefficients in terms of a transmission 2x2 matrix $\mathbf{T}$  for TE:

\begin{equation}
T_{1,n} \begin{pmatrix}
c_n \\
d_n 

\end{pmatrix}
=
T_{2,n} \begin{pmatrix}
a_n \\
b_n \\
\end{pmatrix}
\label{eq:33}
\end{equation}

where $T_{\textit{(1,2}),n}$ is given as:

\begin{equation} \label{Mx}
T_{\textit{i,n}} = 
\begin{pmatrix}
H^{(1)}(k_i \rho) & H^{(2)}(k_i \rho) \\
\frac{\omega \epsilon_i}{\beta_i} {H^{(1)}}'(k_i \rho) & \frac{\omega \epsilon_i}{\beta_i} {H^{(2)}}'(k_i \rho)

\end{pmatrix}
\end{equation} 

where the primed version of the Hankel function $H^{(1,2)}$ indicates its derivative.
Between \((c_n, d_n)\) and \((a_{n+1}, b_{n+1})\) we have:

\begin{equation}
T_{2,n+1} \begin{pmatrix}
c_{n} \\
d_{n} 

\end{pmatrix}
=
T_{1,n+1} \begin{pmatrix}
a_{n+1} \\
a_{n+1} \\
\end{pmatrix}
\label{eq:33}
\end{equation}

Eliminating the coefficients $c_n,d_n$, we obtain a relationship between adjacent pairs of cladding periodic cells:

\begin{equation}
\begin{pmatrix}
a_{n+1} \\
b_{n+1}
\end{pmatrix} = 
\begin{pmatrix}
W & X^{*} \\
X^* & W
\end{pmatrix}
\begin{pmatrix}
a_n \\
b_n
\end{pmatrix}
\label{eq:an12}
\end{equation}

where

\begin{equation}
\begin{aligned}
    W &= P_{21}P_{12} + Q_{21}Q^*_{12} \\
    X &= P_{21}Q_{12} + Q_{21}P^*_{12} \\
    \\
    P_{12} &= 
    \frac{-j k_2 \pi \rho}{4} \left[ H^{(1)}(k_{\rho 2} \rho) {H^{(2)}}'(k_{\rho 1} \rho) - H^{(1)}(k_{\rho 1} \rho) {H^{(2)}}'(k_{\rho 2} \rho) \right] \\
    Q_{12} &= 
    \frac{-j k_2 \pi \rho}{4} \left[ H^{(1)}(k_{\rho 2} \rho) {H^{(1)}}'(k_{\rho 1} \rho) - H^{(1)}(k_{\rho 1} \rho) {H^{(1)}}'(k_{\rho 2} \rho) \right]
\end{aligned}
\label{eq:W_TE}
\end{equation}

From Bloch's theorem, we can also write the eigenvalue equation as:

\begin{equation}
\begin{pmatrix}
a_{n+1} \\
b_{n+1}
\end{pmatrix} = 
exp(-jK\Lambda)T_{\textit{i,n}}{T^{-1}_{\textit{i,n+1}}}
\begin{pmatrix}
a_n \\
b_n
\end{pmatrix}
\label{eq:matrix_eq2}
\end{equation}

The relation in Eq. (\ref{eq:matrix_eq2}) is compared to Eq. (\ref{eq:an12}) and we arrive at the following eigenvalue equation:

\begin{equation}
\begin{pmatrix}
W & Y \\
Y^* & W^*
\end{pmatrix}
\begin{pmatrix}
a_n \\
b_n
\end{pmatrix} = 
R_i exp(-jK\Lambda)T_{\textit{i,n}}{T^{-1}_{\textit{i,n+1}}}
\begin{pmatrix}
a_n \\
b_n
\end{pmatrix}
\label{eq:eiv}
\end{equation}

In Eq. (\ref{eq:eiv}), $R_i$ is the ratio $\sqrt{\frac{\rho_{(n-1),2}}{\rho_{n,2}}}$ which indicates the leakage rate along the structure. Additionally, we can write $L(a_n,b_n)=\lambda(a_n,b_n)$ where $L$ is the matrix operator that translates an eigenvector $L(a_n,b_n)$ on the left-hand side onto the right-hand side multiplied by the eigenvalue $\lambda=\exp{(\Lambda \alpha-jK\Lambda)}$ where $\ln R=\Lambda \alpha$. We replace $\alpha-jK$ with $K'$ and solve for $K'$. Using the Chebyshev polynomials for a structure with $N$-periods, and using $A$ and $B$ after matrix inversion of $T^{-1}_n T_{n+1}$, we obtain: 

\begin{equation}
\begin{aligned}
&\begin{pmatrix}
AU_{N-1}(x) - U_{N-2}(x) & BU_{N-1}(x) \\
B^{*}U_{N-1}(x) & A^{*}U_{N-1}(x) - U_{N-2}(x)
\end{pmatrix}
\begin{pmatrix}
a_n \\
b_n
\end{pmatrix} 
= e^{-jNK'\Lambda} 
\begin{pmatrix}
a_n \\
b_n
\end{pmatrix}
\end{aligned}
\label{eq:eiv}
\end{equation}

where \( x = \frac{A + A^*}{2} \) and the Chebyshev polynomials \( U_N(x) \) are given as

\begin{align}
U_N(x) &= \frac{\sin((N+1)\theta)}{\sin(\theta)} \tag{a} \\
x &= \cos(\theta) = \frac{A + A^*}{2} \tag{b}
\end{align}

The eigenvalue \( \lambda = \exp(-jK'\Lambda) \) has two possible values that correspond to the outgoing and incoming waves within the structure.

\begin{equation}
\lambda_{\pm} = \frac{A + A^*}{2} \pm \sqrt{\left( \frac{A + A^*}{2} \right)^2 - 1}
\end{equation}

and the corresponding eigenvectors are:
\[
\begin{pmatrix}
| & & | \\
\mathbf{v}_1 & \vdots & \mathbf{v}_2 \\
| & & |
\end{pmatrix}
\]

where \( \mathbf{v}_1 \) and \( \mathbf{v}_2 \) are

\begin{equation}
\mathbf{v}_1 = \gamma^+_m \begin{pmatrix}
B \\
\lambda_+ - A
\end{pmatrix}, \quad
\mathbf{v}_2 = \gamma^-_m \begin{pmatrix}
\lambda_+ - A^* \\
B
\end{pmatrix}
\end{equation}

where \( \gamma^{\pm}_m \) refers to the matching coefficient between layers with \( \gamma^+_1 \) being the adjustment coefficient between the core (source region) and the first cladding layer (\( a_{1,1} \) and \( a_{1,2} \)).
The amplitude coefficients in terms of the eigenvalues are:

\begin{equation}
\begin{pmatrix}
a_n \\
b_n
\end{pmatrix} = V \begin{pmatrix}
e^{-j(n-1)K'\Lambda} & 0 \\
0 & e^{-j(n-1)K'\Lambda}
\end{pmatrix} V^{-1} \begin{pmatrix}
\gamma^+_1 \\
\gamma^-_1
\end{pmatrix}
\end{equation}

We will express \( \gamma^{\pm}_1 \) in terms of the field source coefficients (\( E_z \)) placed to the left and right sides of the periodic structure. The matching matrix is given as:

\begin{equation}
\begin{pmatrix}
\gamma^+_1 \\
\gamma^-_1
\end{pmatrix} = V^{-1} T^{-1}_{0,1} \begin{pmatrix}
2a_0 J_0(k_{\rho} \rho_0) \\
-2b_0 J_0(k_{\rho} \rho_0)
\end{pmatrix}
\end{equation}

where the matrix \( T^{-1}_{0,1} \) can be obtained by evaluating Eq. (\ref{Mx}). To evaluate the transmitted field in air, we add the space harmonics contributions to the expression $E_{air}$ evaluated in \ref{seC}. Thus:

\begin{equation}
    E_{2}(\mathbf{R})=E_{air}+E_n
\end{equation}

Where $E_n$ denotes the space harmonics evaluated by the following expression

\begin{equation}
\begin{split}
& E_{n}{(\mathbf{R})} =  \frac{T^{TE}_1}{\Lambda W} 2k_1 
 \sum_{-\infty}^{\infty} \left[ \frac{k_{2zn}}{k_{1zn}} M_{01k_{\rho}}(k_{1zn}) \left( e^{jk_{2zn}z_0} + R^{TE}_1 e^{-jk_{2zn}z_0} \right) \right. \\
& \left. + j \frac{T^{TM}_1}{\Lambda W'} \left(\frac{k_{\text{si}}}{k_{\text{air}}}\right) {N}_{e1k_{\rho}}(k_{1zn}) 
 \left( e^{-jk_{2zn}z_0} - R^{TM}_2 e^{-j2\phi_2}e^{-jk_{2zn}z_0} \right) \right] 
\end{split} \label{E_{2T}}
\end{equation}

$k_{zn}$ and $k_{\rho n}$ are space harmonics related by 

\begin{equation}
    k_{\rho n}=k_{\rho }+\frac{2\pi n}{\Lambda} \quad
    k^2_{zn}=k^2-k^2_{\rho n}
\end{equation}

For backward leaky-wave radiation, Eq. (\ref{E_{2T}}) would be evaluated for the harmonic $n=-1$ space.

\subsection{Surface Impedance}
The impedance of the periodic structure along $\rho$ is evaluated as a sum of the reference impedance ($Z_{ref}$) and the impedance due to the modulation of the surface by the periodic structure. Thus:

\begin{equation} \label{Zs}
Z = j \left[ Z_{ref} + m \cdot \text{Re}(E_{2T} E^*_{\text{inc}}) \right] 
\end{equation} 

Where $E_{inc}$ is the incident wave from the silicon layer and the parameter $m$ is the modulation index.  The second part of Eq. (\ref{Zs}) is related to the power along the periodic surface whereas the first part is derived using the Transverse Resonance Technique (TRT) as explained in the next section.

\begin{comment}

\end{comment}

\begin{figure*}[ht!]
    % First row of three figures
 %   \vspace*{40pt}
    \centering
 %   \begin{subfigure}[b]{1\textwidth}
  %      \includegraphics[width=\textwidth]{Half_Tx.pdf}
   %     \caption{}
   %     \label{fig:5a}
   % \end{subfigure}%
 %   \hfill % spacing between the subfigures
 \vspace*{-1pt} 
    % Second row of two figures
    \begin{subfigure}{1\textwidth}
        \includegraphics[width=\textwidth]{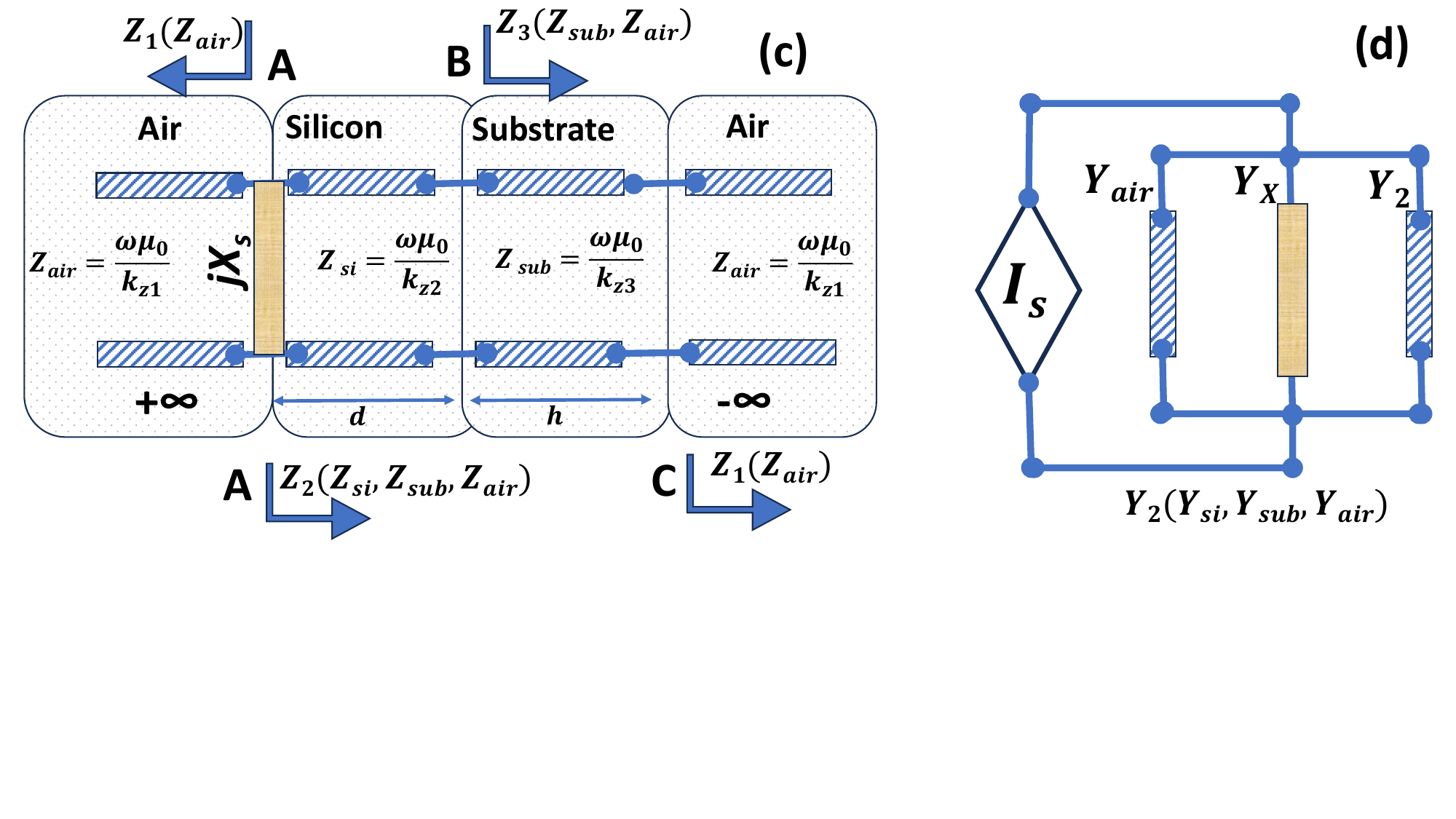}
      %  \caption{}
        \label{fulltx}
    \end{subfigure}%
   % \hfill % spacing between the subfigures  
    \vspace*{-90pt} 
    \caption{Transverse resonance model of a 3 layer Si structures: a) Silicon slab loaded with metasurface and substrate, b) TRM model of the loaded Silicon structure}
    \label{fig:TRMn}
\end{figure*}

The TRT method is powerful in the analysis of dielectric structures at microwave and optical frequencies. In principle, the impedance on a transmission line looking in one direction is equal to the impedance on the same line at the same point but looking in the opposite direction, only separated by a sign ($\pm$). Similar to an RLC circuit, the technique yields the resonant frequencies of layered structures. A transmission line models is shown in Fig. 6.  The model represents the dielectric materials in the stack as lumped elements. The  material impedances are written in terms of their permeability $\mu$ and permittivity $\epsilon$, as well as the respective wave numbers $k_{z1}$, $k_{z2}$, and $k_{z3}$ for air, silicon, and substrate, respectively. Specifically, the impedance for each layer is given by $Z_{\text{medium}} = \frac{\omega \mu_0}{k_{z\text{medium}}}$, where $\omega$ is the angular frequency and $\mu_0$ is the permeability of free space.
 The impedance at the Silicon-Air interface is $Z_{air}=Z_0$. Positioning the observer at A in the absence of the substrate, we can deduce a relationship between $Z_{air}$ and $Z_{si}$ from the transmission line equation as follows.

\begin{align}
    Z^{\rightarrow}_\text{right} &= - Z^{\leftarrow}_\text{left} \\
    Z^{\leftarrow}_\text{left} &= Z_0 \\
    Z^{\rightarrow}_\text{right} &= Z_{\text{si}} \left[\frac{Z_L + jZ_\text{si}\tan(k_{z2}d)}{Z_{\text{si}} + jZ_L \tan(k_{z2}d)} \right] 
 \end{align}

where $Z_L$ is the impedance at the silicon air interface looking into the open circuit of free space, simply $Z_0$=$\frac{\omega\mu}{k_{z0}}$. The transcendental equation after applying the TRM technique yields:

\begin{subequations}\label{eq:MTRM}
    \begin{align}
        \frac{-2Z_0 Z_{\text{si}}}{j(Z_0^2 + Z_{\text{si}}^2)} &= \tan(k_{x2}d) \label{TRM1a} \\
        -2 &= Y_T \tan(k_{x2}d) \label{TRM1b}
    \end{align}
\end{subequations}

where $Y_T = \frac{1}{Z_0} + \frac{1}{Z_{\text{si}}}$ in Eq.~\ref{TRM1b}. The dispersion relation is obtained by simultaneously solving Eq.~\ref{eq:MTRM} and the separation relation $k_{\rho}^2 + k_z^2 = nk_0^2$. When the structure is loaded with a metasurface, we repeat the procedure to obtain the impedance at point A where the admittance $Y_{A,B,C}$ is replaced with $Y_{Ln}$ for brevity. The admittance at point A is related as:

\begin{equation}
Y_{Ln}+Y_0+Y_s=0
\label{Yms}
\end{equation}

where $Y_{Ln}$ is synthesized as:

\begin{subequations}\label{Ylm}
    \begin{align}
        Y_{Ln} &= \left[\frac{Y_{\text{su}}Y_{\text{p}} + jY_{\text{si}}Y_{\text{q}}\tan(k_{z2}d)}{Y_{\text{si}}Y_{\text{q}} + jY_{\text{su}}Y_{\text{p}} \tan(k_{z2}d)} \right] \label{eq:Y_Ln} \\
        Y_{q} &= Y_{\text{0}}Y_{\text{si}} + Y_{\text{si}}Y_{\text{su}}\tan(k_{z2}h) \label{eq:Y_q} \\
        Y_{p} &= Y_{\text{si}}Y_{\text{su}} + Y_{\text{0}}Y_{\text{si}}\tan(k_{z2}h) \label{eq:Y_p}
    \end{align}
\end{subequations}

Eq. \ref{Ylm} can be simplified if we consider the conditions of a grounded slab at point B in which the impedance between the silicon and the substrate is zero. This induces a short circuit at point B and the TRM equation reduces to:
\begin{equation}
    Y_{\text{0}} + Y_s - jY_{\text{si}} \cot(k_{z1}d) = 0 \label{eq:transverse_resonance}
\end{equation}

Using the small-angle approximation for the cotangent function, $\cot(\theta) \approx \frac{1}{\theta}$ allows us to extract the real and imaginary parts separately as:
\begin{equation}
    Y_{\text{0}}d + Y_s d - jY_{si} \frac{1}{(\beta - j\alpha)} = 0 \label{eq:small_angle_approx}
\end{equation}

With the assumption that $Y_s$ is purely reactive, it can be represented as $Y_s = jX_s$, where $X_s$ is the sheet impedance. Substituting into Equation \eqref{eq:small_angle_approx}, and separating into real and imaginary parts we get:

\begin{align}
    \alpha d Y_{\text{0}} + d X_s \beta - Y_{\text{si}} &= 0 \label{eq:real_part} \\
    \beta d Y_{\text{0}} + d X_s \alpha &= 0 \label{eq:imaginary_part}
\end{align}

Solving the system composed of Equations \eqref{eq:real_part} and \eqref{eq:imaginary_part} for $X_s$ and $d$, we find:
\begin{align}
    X_s &= -\frac{Y_{\text{air}} \beta}{\alpha} \label{eq:X_s} \\
    d &= \frac{Y_{\text{si}} \alpha}{Y_{\text{0}} (\alpha^2 - \beta^2)} \label{eq:h}
\end{align}

Equations \eqref{eq:X_s} and \eqref{eq:h} provide the expressions for the sheet impedance $X_s$ and the height $d$ in terms of the wave parameters $\alpha$, $\beta$, the air admittance $Y_{\text{air}}$, and the characteristic admittance $Y_{\text{si}}$.

\section{Design Considerations and Results}

In this section, we explore two distinct designs operating in the terahertz and optical ranges, with wavelengths of 14$\mu$m and 1.5$\mu$m, respectively. We derive the electric fields in free space i.e. above the silicon surface using a combination of integral equations in the discretized unit radiators. The phase modulation $(m)$ on the metasurface is realized by using unit cells of identical heights but different lengths to vary the refractive indices. 

\begin{figure*}[t]
    \centering
    % First subfigure
    \begin{subfigure}[b]{0.4\textwidth}
        \includegraphics[width=\textwidth]{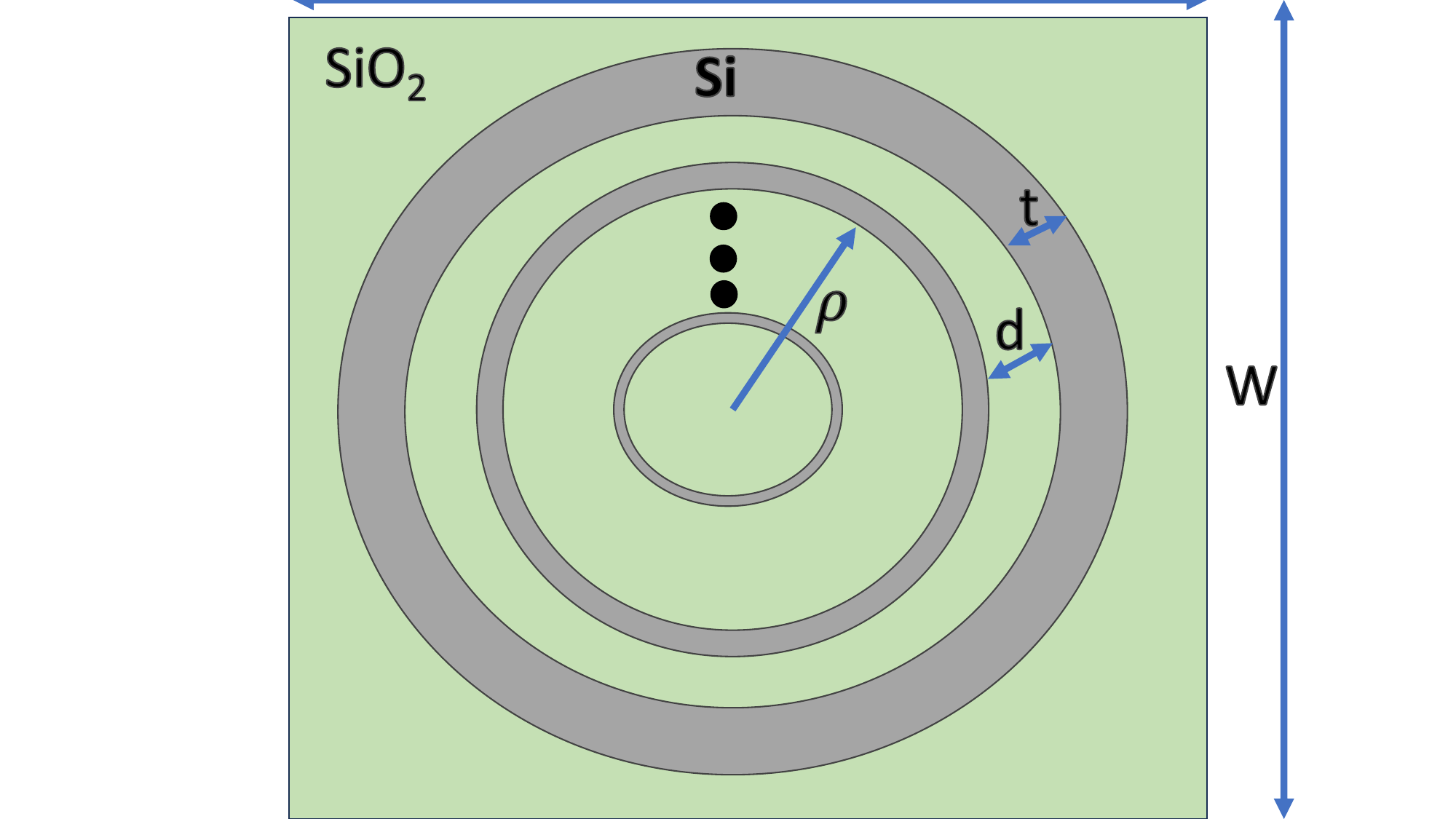}
        \caption{}
        \label{fig:sub1}
    \end{subfigure}%
    \hfill % spacing between the subfigures
    % Second subfigure
    \begin{subfigure}[b]{0.6\textwidth}
        \includegraphics[width=\textwidth]{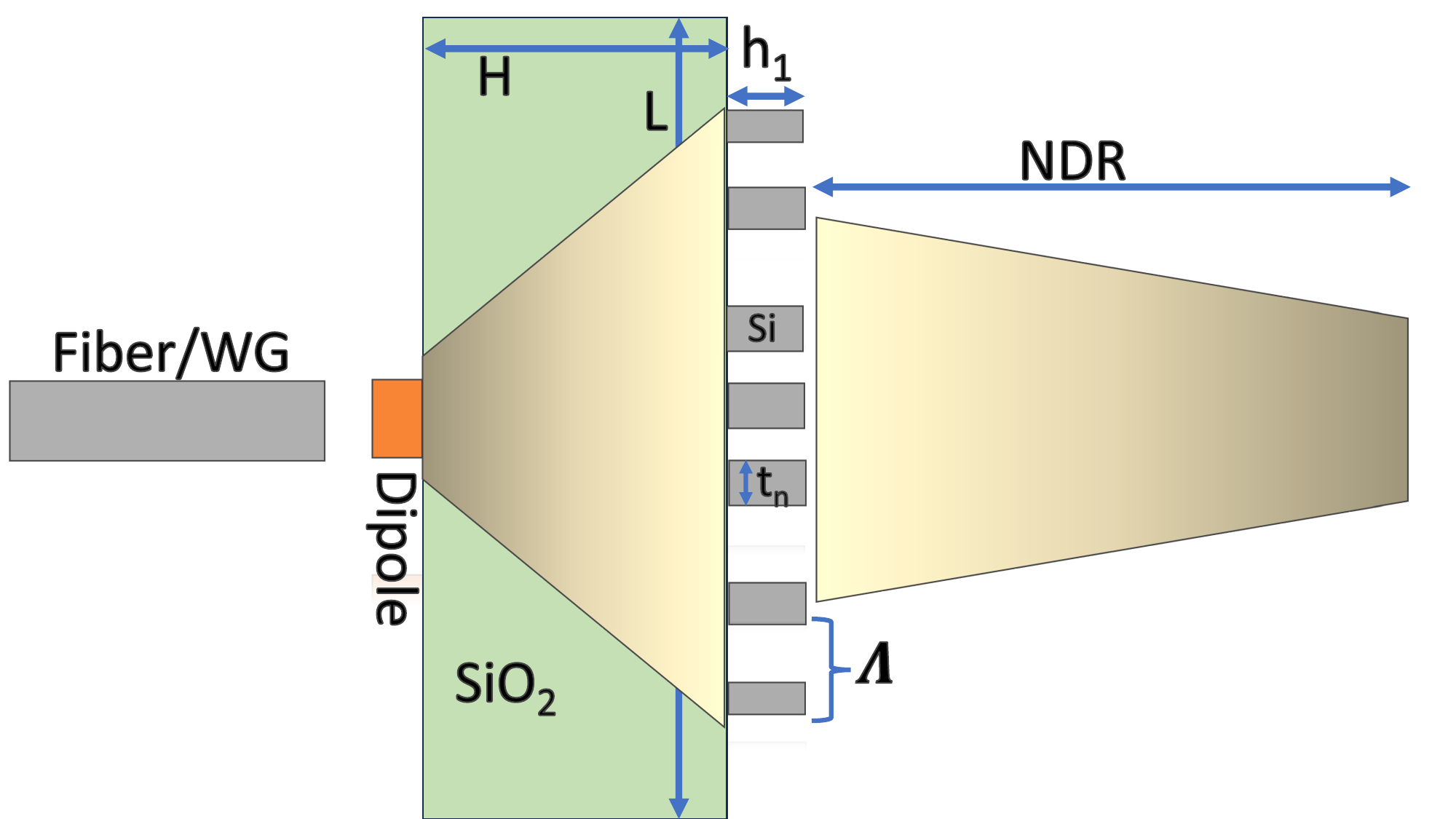}
        \caption{}
        \label{disc_1550}
    \end{subfigure}
    \caption{Bessel Lens structure: a) 2D cross section with concentric rings on top of SiO$_2$. b)Longitudinal cross-section with a dipole source at the input and the Bessel beam launched at the output}
    \label{GenBL}
\end{figure*}

In Fig.~\ref{GenBL} an external source launches EM waves at 1550 nm, with a linear polarizer simulating a linearized dipole moment. Alternatively, a linearly polarized dipole is excited at THz frequencies. A secondary wave is generated in the SiO$_2$ medium of dimensions LxWxH. A modulating surface of silicon rings is introduced with thickness $\boldsymbol{t}$, separated by distance $\boldsymbol{d}$ on top of the SiO$_2$ substrate. The total metalens radius is $\rho$ and encompasses all the rings from the center to the edge of the launcher. The EM fields across the scattering metasurface are calculated using integral equations for self-consistent field ($\mathbf{E}$, $\mathbf{H}$) computation.

We will denote the field in free space from the unmodulated silicon structure as $\mathbf{E}_{ref}$ when the structure is excited by the current $-j\omega\mu_0\mathbf{J}(\mathbf{R}-\mathbf{R'})$. When the same current is used to excite the modified structure in the presence of an impedance metasurface, the resultant electric-field integral equation can be written as:

\begin{align} \label{IE2}
\mathbf{E}(\mathbf{R}) &= \mathbf{E}_{\text{ref}}(\mathbf{R})+ 
 \int \left[ k^2(\mathbf{R'}) - k_{\text{ref}}^2(\mathbf{R'}) \right] \mathbf{G}(\mathbf{R}, \mathbf{R'}) \mathbf{E}(\mathbf{R'}) \, dV',
\end{align}

where $[k(\mathbf{R'}) - k_{\text{ref}}\mathbf{(R')}]$ =$k_0\sqrt{\epsilon(\mathbf{R'})} - k_0\sqrt{\epsilon_{\text{ref}}(\mathbf{R'})}$.

The electric field in the $i^{th}$ ring can be decomposed vectorially in terms of the cylindrical coordinate system:

\begin{equation}
\mathbf{E}^i(\mathbf{R}) = \left( \mathbf{{E}}_{\rho}^{(i)} + \mathbf{{E}}_{\phi}^{(i)} + \mathbf{{E}}_{z}^{(i)}  \right)
\end{equation}
Each component in $\mathbf{E}^i(\mathbf{R})$ is calculated using the integral equation in Eq. \ref{IE2}. The integral equation also accounts for the interaction between unit cells along the scattering cross-section. The coupling relationship between the Dyadic elements $(\rho,\phi,z)$ leads the Green's function in Eq.\ref{IE2} with $N$ elements. We denote a matrix \( \mathbf{A}_i \) to encapsulate that coupling between the $i^{th}$ and $j^{th}$ elements.

\begin{subequations}
\begin{equation} \label{Ac}
\mathbf{A}_{i\rightarrow{j}} = 
\begin{bmatrix}
G_{\rho\rho} & G_{\rho\phi} & G_{\rho z} \\  % Corrected A to G
G_{\phi\rho} & G_{\phi\phi} & G_{\phi z} \\  % Corrected A to G
G_{z\rho} & G_{z\phi} & G_{zz}
\end{bmatrix}
\end{equation}
\begin{equation} \label{Gc}
\mathbf{G}_{i\rightarrow{j}} = \iint_{\Omega} \mathbf{A}_{i\rightarrow{j}}(\rho_i,z_i|\rho',z')\rho'd\rho'dz'
\end{equation}
\end{subequations}

 The final integral equation describing the electric field in the structure can be written as:

\begin{subequations}
    \label{eq:GE1}
    \begin{equation}
        \begin{bmatrix}
            \mathbf{E}^{(n)}_1 \\
            \mathbf{E}^{(n)}_2 \\
            \vdots \\
            \mathbf{E}^{(n)}_N
        \end{bmatrix} =  [\left( I - \mathbf{A} \Delta\epsilon / k_0 \right) ]^{-1}\begin{bmatrix}
            \mathbf{E}^{(n)}_{0,1} \\
            \mathbf{E}^{(n)}_{0,2} \\
            \vdots \\
            \mathbf{E}^{(n)}_{0,N}
        \end{bmatrix}
    \end{equation}
    \label{eq:GE2}
    \begin{equation}
        \Delta\epsilon = \begin{bmatrix}
            \mathbf{I} (\epsilon_1 - \epsilon_{\text{ref},1}) & \mathbf{0} & \cdots & \mathbf{0} \\
            \mathbf{0} & \mathbf{I} (\epsilon_2 - \epsilon_{\text{ref},2}) & \cdots & \mathbf{0}_3 \\
            \vdots & \vdots & \ddots & \vdots \\
            \mathbf{0} & \mathbf{0} & \cdots & \mathbf{I} (\epsilon_N - \epsilon_{\text{ref},N})
        \end{bmatrix}
    \end{equation}
\end{subequations}

$\mathbf{E}^{(n)}_{0,n}$ is the field in the reference structure in the nth element, and the matrix A is an N by N matrix representing the interaction between n-1 elements in the scattering cross-section:

\[
A = \left[
\begin{array}{ccc}
    A_{\leftrightarrow_{i=1,j=1}}^{11} & \cdots & A_{\leftrightarrow_{i=1,j=N}}^{1N} \\
    \vdots & \ddots & \vdots \\
    A_{\leftrightarrow_{i=N,j=1}}^{N1} & \cdots & A_{\leftrightarrow_{i=N,j=N}}^{NN}
\end{array}
\right] = \lVert A_{\leftrightarrow_{ij}} \rVert^N
\]

\subsection{Case I: THz Regime} \label{seCD}

At a wavelength of 14 $\mu$m, the geometric parameters of the rectangular silicon box are defined as \( L \times W \times H = 450 \, \mu\text{m} \times 450 \, \mu\text{m} \times 100 \, \text{nm} \). A configuration of 12 concentric rings, each exhibiting distinct thicknesses, is positioned atop the silicon substrate. The total radius (\( \rho \)) of the metasurface lens is quantified as 300 $\mu$m, and each constituent unit cell has a distinctive refractive index tailored to match the phase of the Bessel function along the radius \( \rho \) of the Bessel launcher.

The implementation of [Eq.(\ref{IE2})-Eq.(\ref{eq:GE1})] is compared to the simulation of the structure in the CST microwave studio software and the results are shown in Fig. 8. The integral equation solver in CST is used. An asymmetric line source is also used to emulate a linearly polarized incident wave. The simulation plots are compared to the simulations from CST and they are shown to be in agreement. The lens shows power focus along the z-direction. The thicknesses of the unit ring cells in the Bessel lens in this case range from $300\mu{m}$ to $50\mu{m}$, referencing Fig.~\ref{GenBL}. 
The process of discretization is conducted on a per-ring basis. For a cylindrical radius of 300$\mu$m, we implement three distinct sets of discretization. The cells encompassed within the initial 150$\mu$m are discretized into six elements per ring, aligned along the lens radius. The secondary set of discretization is executed between 175$\mu$m and 230$\mu$m, with the final two rings being discretized utilizing two elements each. As the thickness of the ring diminishes with increasing distance from the center, the power leakage into free space increases, proportionally related to the value of $\alpha\approx\sqrt{\frac{1}{\rho}}$.

\begin{figure*}[ht] \label{88}
    % First row of three figures
    \vspace*{40pt}
    \centering
    \begin{subfigure}[b]{0.33\textwidth}
        \includegraphics[width=0.9\textwidth, height=3cm]{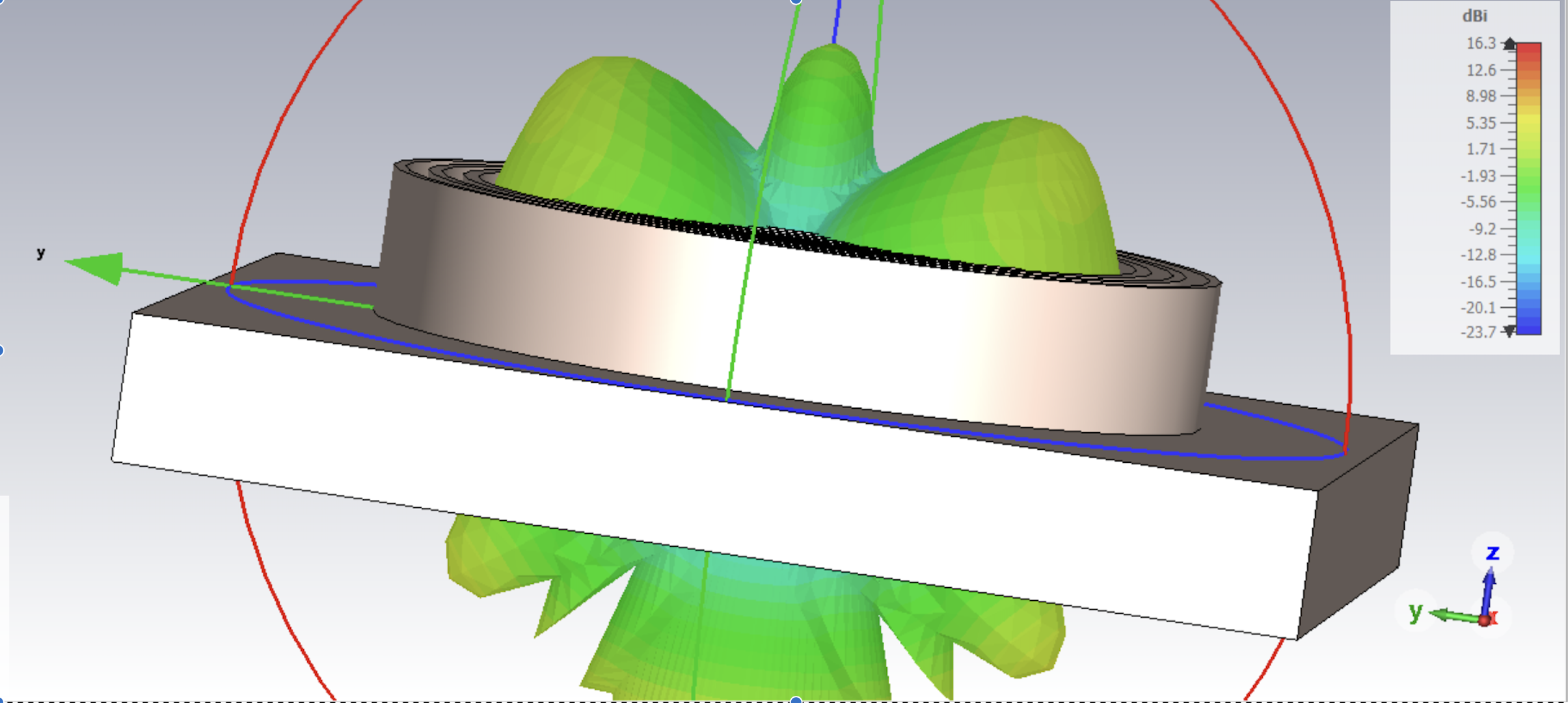}
        \caption{}
        \label{fig:5a}
    \end{subfigure}%
    \hfill % spacing between the subfigures
    \begin{subfigure}[b]{0.33\textwidth}
        \includegraphics[width=0.9\textwidth, height=3cm]{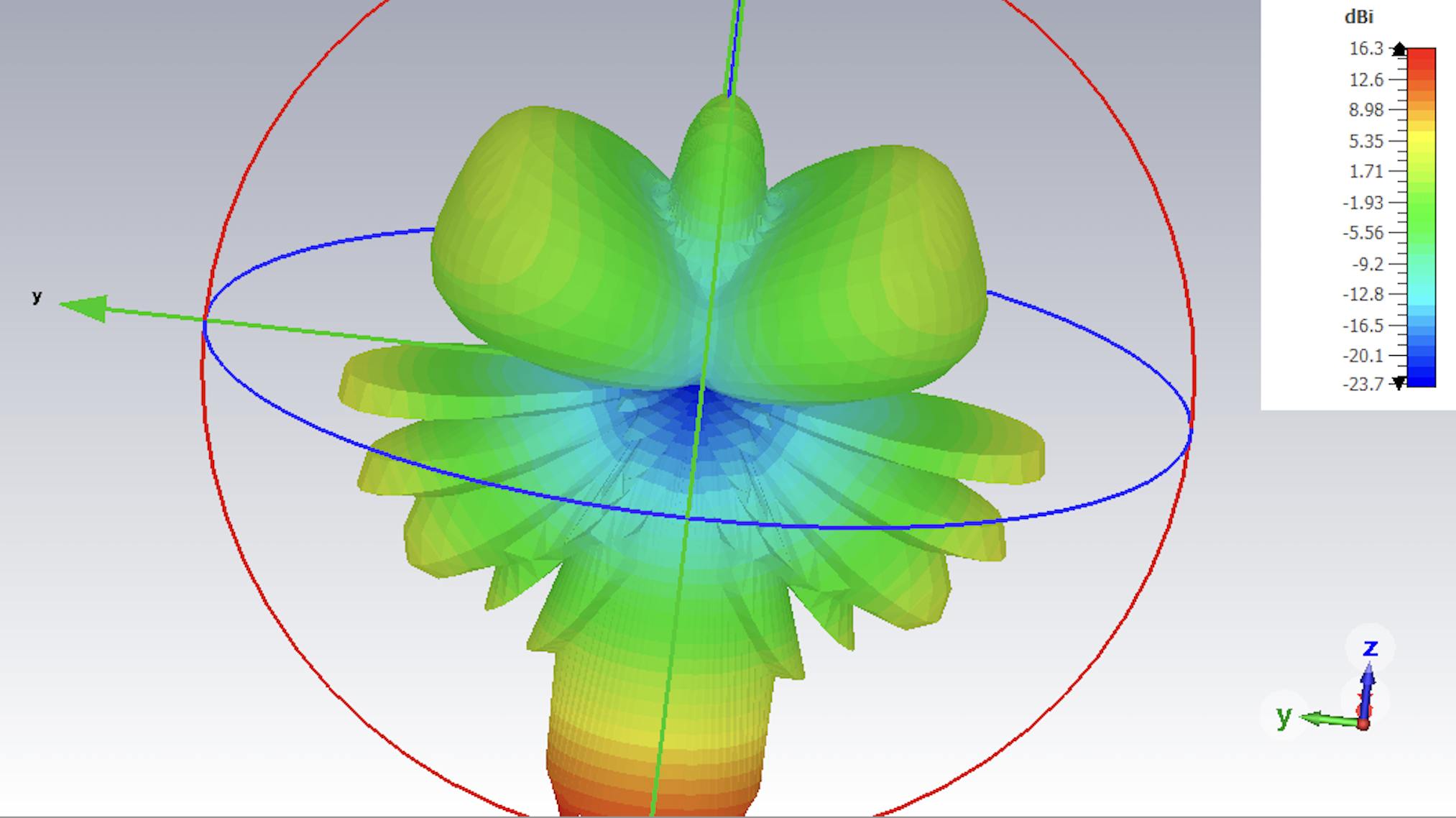}
        \caption{}
        \label{fig:5b}
    \end{subfigure}%
    \hfill % spacing between the subfigures
    \begin{subfigure}[b]{0.33\textwidth}
        \includegraphics[width=0.9\textwidth, height=3cm]{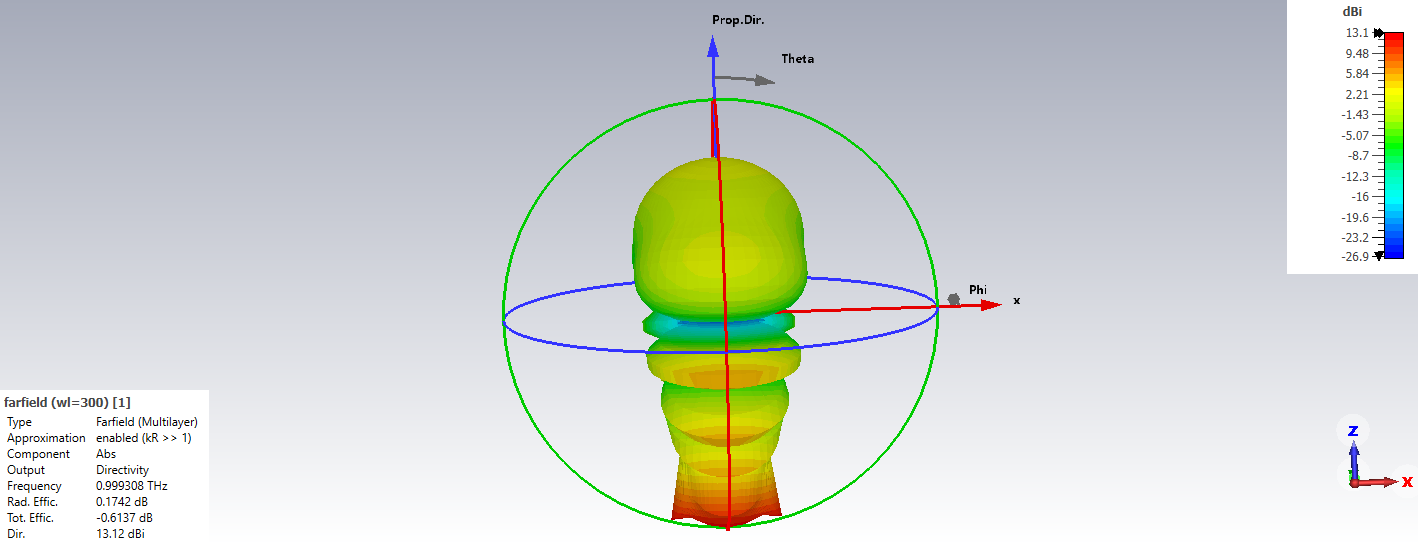}
        \caption{}
        \label{fig:5c}
    \end{subfigure}
    
    % Second row of two figures
    \begin{subfigure}[b]{0.5\textwidth}
        \includegraphics[width=\textwidth]{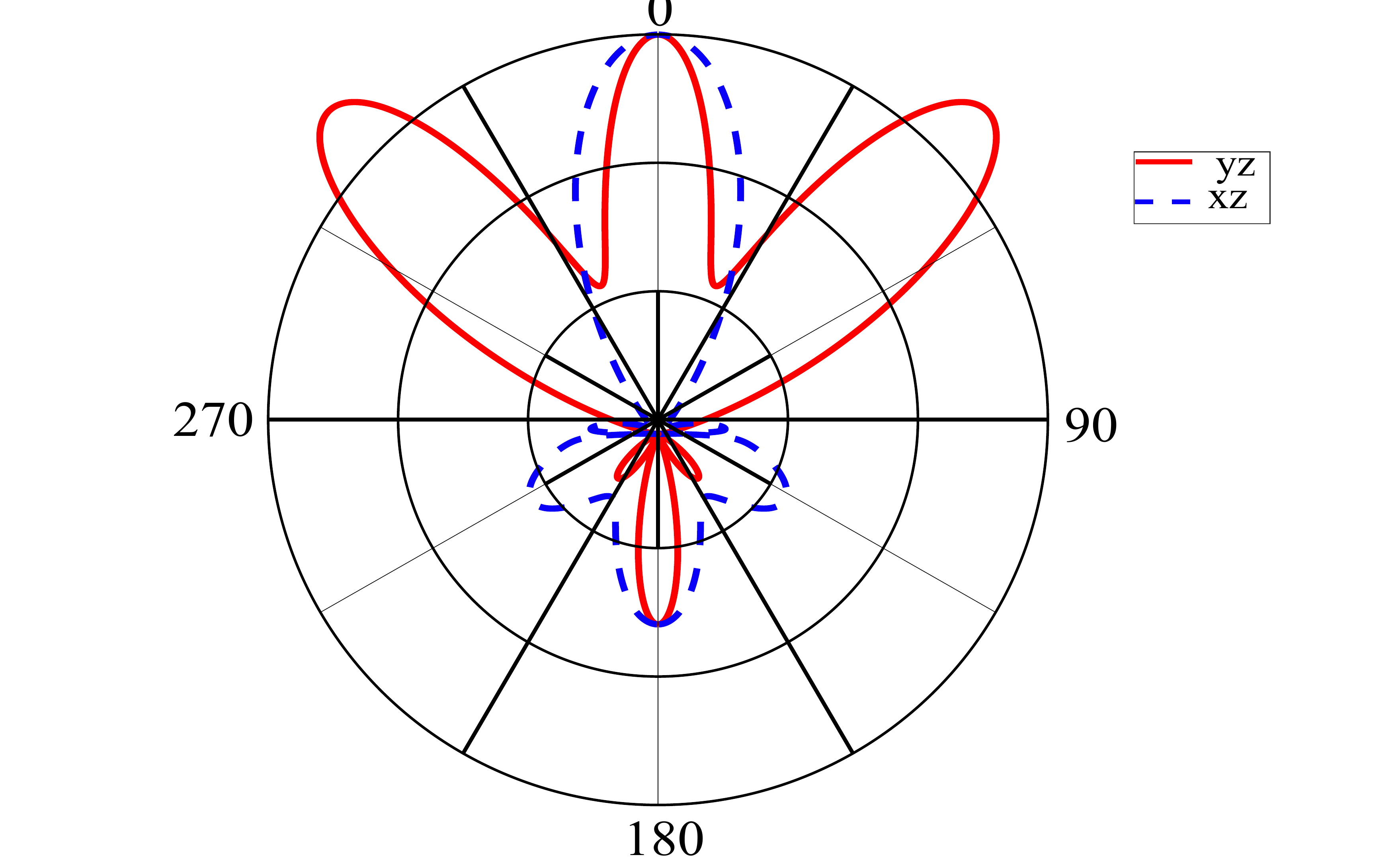}
        \caption{}
        \label{polar_14THz}
    \end{subfigure}%
    \hfill % spacing between the subfigures
    \begin{subfigure}[b]{0.5\textwidth}
        \includegraphics[width=\textwidth]{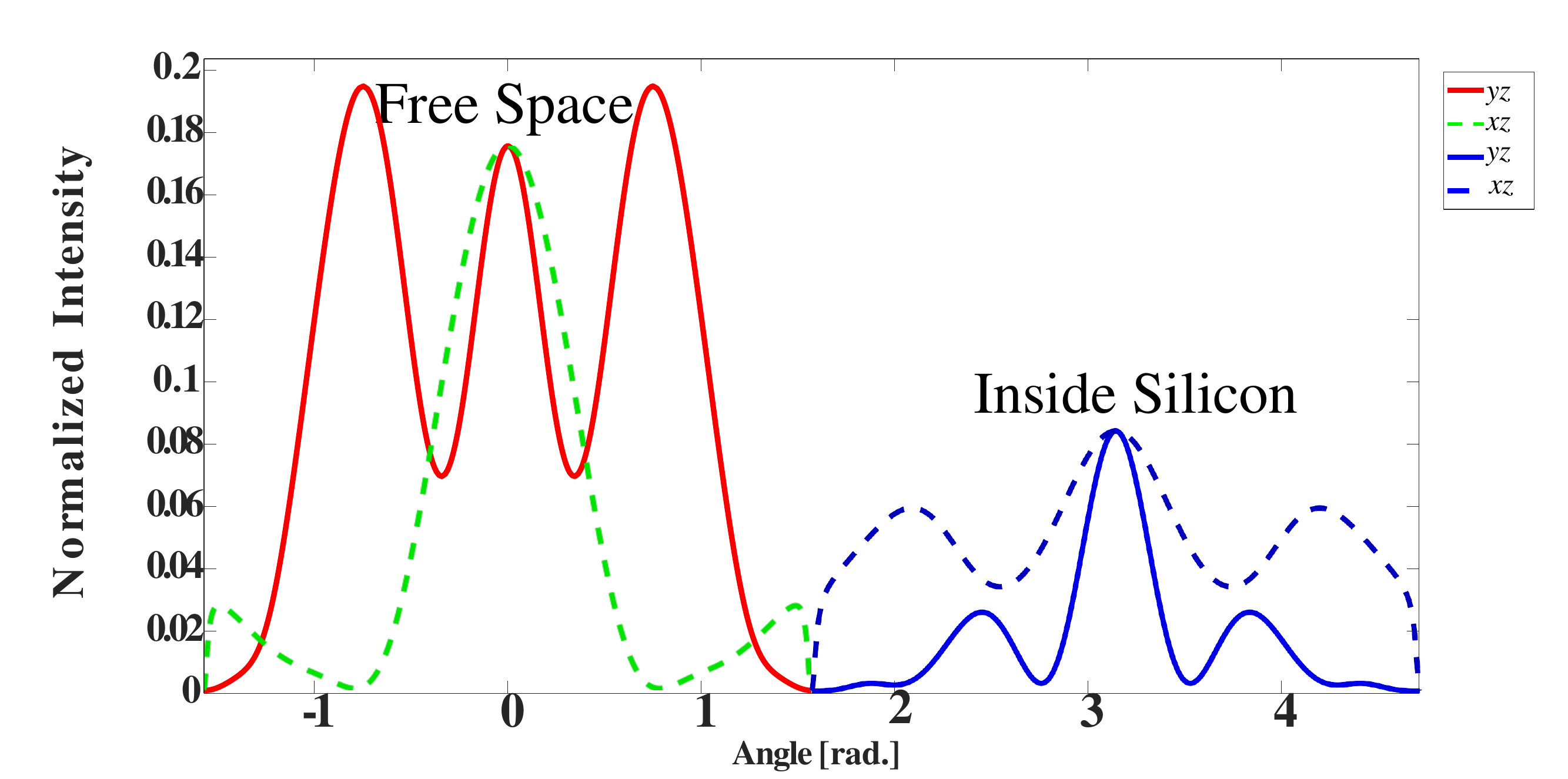}
        \caption{}
        \label{fig:5e}
    \end{subfigure}
    
    \caption{Farfield radiation patterns comparing CST studio and Green's Function Modelling. a) Y-Z radiation in CST Simulation with structure, b)Y-Z radiation in CST Simulation without structure, c) X-Z radiation in CST, d) Polar plots from the integral equation above the metasurface of the lens: Y-Z and X-Z farfields. e) Normalized power intensity of the radiated field}
    \label{fig:5}
\end{figure*}

In Fig. (\ref{fig:5e}), we show the electric field distribution along the structure both inside and outside the cylindrical structure. The region denoted as "Inside silicon" represents the fields inside the bounded silicon region. The intensity of the fields inside is about 3dB lower than the emitted radiation. The amount of energy confined inside the structure depends on the value of $\alpha$, which is proportional to the attenuation constant along $\rho$. A small value of $\alpha$ would increase the radiation efficiency of the lens, with narrower beam width. However, this would require a metasurface with a bigger ring radius. The presence of a metasurface at a frequency of 2 THz, results in a significant boost in forward emission compared to other directions.

\subsection{Case II: Optical Regime} \label{seCD}
The design is repeated at optical frequencies i.e. 1.5 $\mu$m and the lens is again designed to direct power along the \( z \)-axis like a Bessel beam in the C optical band. As in the THz case, the discretization is scaled such that the widths of the unit cells are optimized based on the value of $\beta$ and $\alpha$ at 1550nm with unit cell height fixed to 100nm. The thicknesses range from \( 2.5\mu \)m down to \( 0.5\mu \)m. 

\begin{figure*}[h!] 
%\vspace*{100pt}
\label{beam_prop}
    \centering
    % First subfigure
    \begin{subfigure}[b]{0.5\textwidth}
        \centering
        \includegraphics[width=\textwidth]{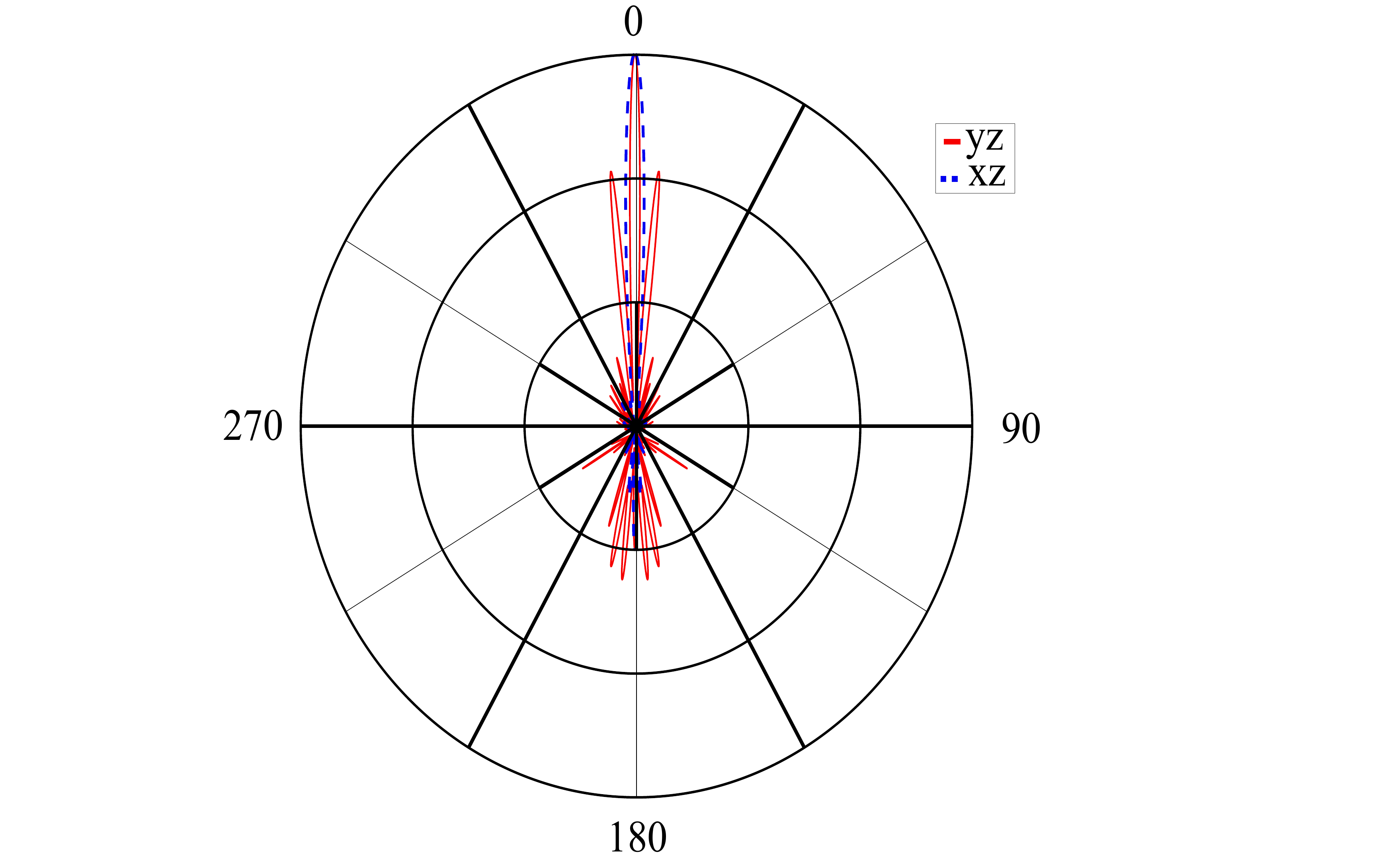}
        \caption{}
        \label{fig:sub1}
    \end{subfigure}%
    \hfill % spacing between the subfigures
    % Second subfigure
    \begin{subfigure}[b]{0.5\textwidth}
        \centering
        \includegraphics[width=\textwidth]{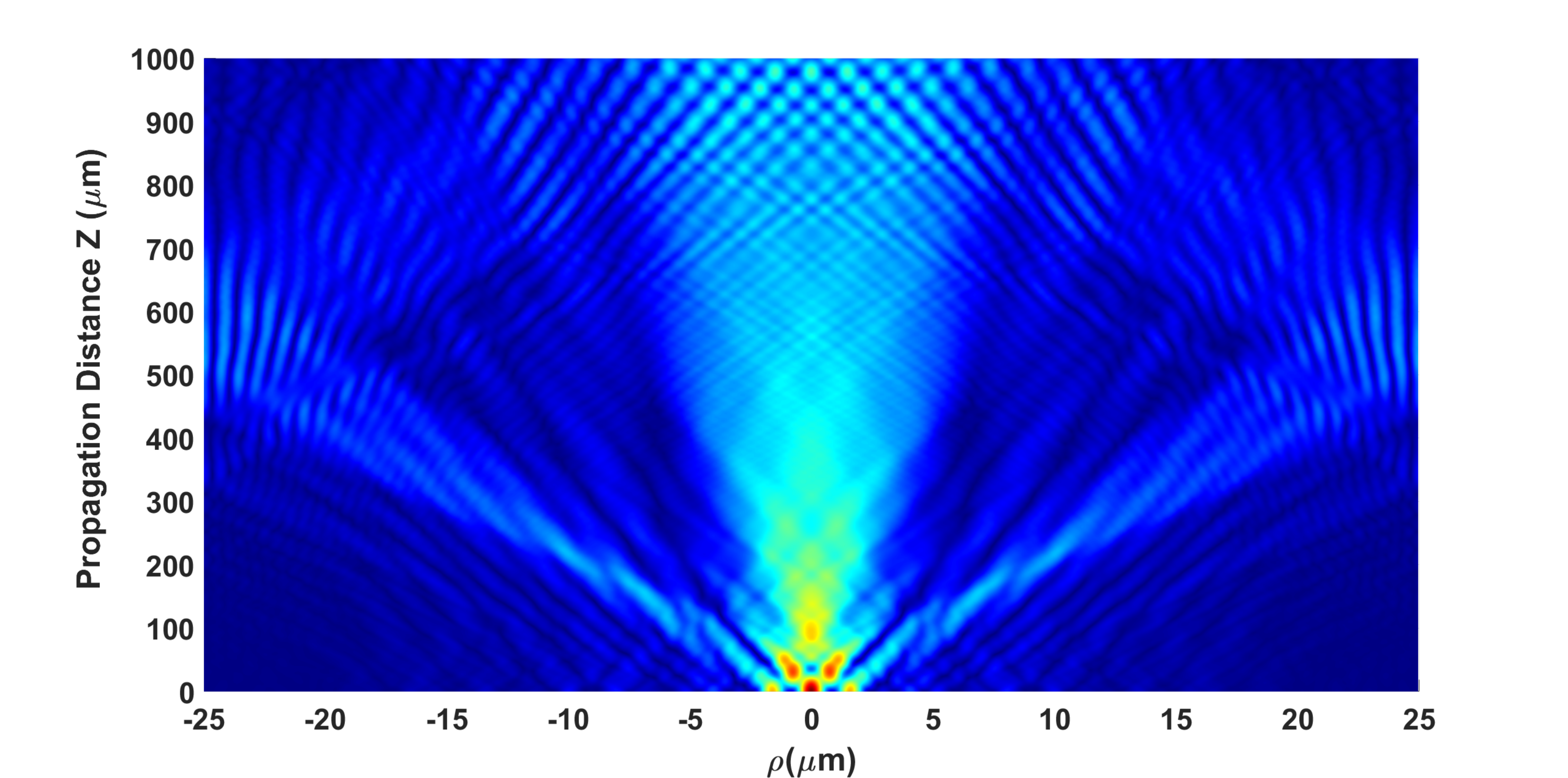}
        \caption{}
        \label{disc_1550}
    \end{subfigure}
    % New line for the third subfigure
    \caption{Analytical Results at 1550nm: (a) Polar plot at 1550nm for XZ and YZ planes, (b) Beam Propagation in free space for the discretized lens}
    \label{bessel_1550}
\end{figure*}
The normalized electric field along the scattering cross section is used to produce the beam propagation along the z-axis as shown in Fig.\ref{disc_1550}. The figure shows a sustained non-diffracting range of approximately 500$\mu$m.

\section{conclusion}

%In this study, we have established a comprehensive theoretical framework that facilitates the design of devices capable of generating Bessel beams within silicon waveguide infrastructures. Our approach is grounded in the development and application of various mathematical instruments, including vector wave functions. These functions are crucial for the eigenmode decomposition of waves within homogeneous environments, enabling a precise analysis of wave behavior in such settings.

%Furthermore, we have derived the Dyadic Green's functions for both electric and magnetic fields. These functions play a pivotal role in simulating the electric field distributions within and external to bounded media, thereby providing the field dynamics across different regions. In addition, our study utilizes holographic methods to model the desired metasurface impedance needed to produce the desired Bessel beam characteristics.
%We demonstrate the effectiveness of our technique through the generation of far-field patterns of Bessel waves at optical frequencies. The metasurface patterns appear as arrays of concentric rings situated atop a rectangular silicon waveguide, which collectively act to create a Bessel beam. This is achieved by maintaining the radiation angle of the beam as one moves away from the center of the concentric rings. This demonstration highlights the potential of the proposed design methodology for advancing the application of Bessel beams in integrated photonics and communication systems. 

This study establishes a comprehensive theoretical framework for designing devices capable of generating Bessel beams within a silicon waveguide platform. The platform integrates metasurface patterns that appear as arrays of concentric rings situated atop a rectangular silicon waveguide, which collectively create a Bessel beam by maintaining the radiation angle as one moves away from the center of the concentric rings.

The approach is grounded in the development and application of various mathematical tools, including vector wave functions, which are crucial for the eigenmode decomposition of waves within homogeneous environments, enabling precise analysis of wave behavior. Furthermore, the study derives Dyadic Green's functions for both electric and magnetic fields. These functions are pivotal in simulating the electric field distributions within and external to bounded media, thereby elucidating field dynamics across different regions. Additionally, holographic methods are employed to model the metasurface impedance required to achieve the desired Bessel beam characteristics.

The effectiveness of this technique is demonstrated through the generation of far-field patterns of Bessel waves at optical and terahertz frequencies. This demonstration underscores the potential of the proposed design methodology to advance the application of Bessel beams in integrated photonics and communication systems.

\section{Appendix A}

\begin{align*}
a & = \frac{R_2 e^{-j2\phi_2}}{W}\left[ \mathbf{M'}{(k_{2z})} + R^{TE}_1 \mathbf{M'}{(-k_{2z})}\right], \\
b &= \frac{R^{TE}_1 }{W}\left[ R^{TE}_2e^{-j2\phi_2}\mathbf{M'}{(k_{2z})} + R^{TE}_1 \mathbf{M'}{(-k_{2z})}\right], \\
c &= \frac{R^{TM}_2 e^{-j2\phi_2}}{W'}\left[ \mathbf{N'}{(k_{2z})} + R^{TM}_1 \mathbf{N'}{(-k_{2z})}\right], \\
d &= \frac{R^{TM}_1 }{W}\left[ R^{TM}_2e^{-j2\phi_2}\mathbf{N'}{(k_{2z})} + R^{TM}_1 \mathbf{N'}{(-k_{2z})}\right], \\
e &= \frac{1+R^{TE}_1 }{W}\left[ R^{TE}_2e^{-j2\phi_2}\mathbf{M'}{(k_{2z})} + \mathbf{M'}{(-k_{2z})}\right], \\
f &= \frac{1+R^{TM}_1 }{W}{\frac{k_{si}}{k_{air}}}\left[ R^{TM}_2e^{-j2\phi_2}\mathbf{N'}{(k_{2z})} + \mathbf{N'}{(-k_{2z})}\right], \\
g &= \frac{1+R^{TE}_3 }{W}e^{-j2({\phi_2}-\phi_1)}\left[ \mathbf{M'}{(k_{2z})} + R^{TE}_3\mathbf{M'}{(-k_{2z})}\right], \\
h &= \frac{1+R^{TM}_3 }{W}e^{-j2({\phi_2}-\phi_1)}{\frac{k_{si}}{k_{sub}}}\left[ \mathbf{N'}{(k_{2z})} + R^{TM}_3\mathbf{N'}{(-k_{2z})}\right].
\end{align*}
where 
\begin{align*}
\phi_1 &= k_{1z}d, \quad \phi_2 = k_{2z}d, \\
k_{1z} &= \left( k_1^2 - k_{x1}^2 \right)^{\frac{1}{2}}, \quad k_{2z} = \left( k_2^2 - k_{x2}^2 \right)^{\frac{1}{2}}, \\
k_1 &= \omega \left( \mu_0 \epsilon_1 \right)^{\frac{1}{2}}, \quad k_2 = \omega \left( \mu_0 \epsilon_2 \right)^{\frac{1}{2}}, \\
R^{TE}_{1,3} &= \frac{k_{2z} - k_{{1,3}z}}{k_{2z} + k_{{1,3}z}}, \\
R^{TM}_{1,3} &= \frac{\left( \frac{k_{1,3}^2}{k_{{1,3}z}} k_{2z} - \frac{k_2^2}{k_{2z}} k_{{1,3}z} \right)}{\left( \frac{k_{1,3}^2}{k_{{1,3}z}} k_{2z} + \frac{k_2^2}{k_{2z}} k_{1z} \right)}, \\
W &= 1 - R^{TE}_1 R^{TE}_2 e^{-j2\phi_2}, \quad W' = 1 - R^{TM}_1 R^{TM}_2 e^{-j2\phi_2}.
\end{align*}

\section{Appendix B}
\begin{equation}
\begin{aligned}
G^{e}_{si}(\mathbf{R},\mathbf{R'}) = -\hat{z} \hat{z} \delta(\mathbf{R} - \mathbf{R}')+& \\
&\hspace{-10em}\iint_{0}^{\infty} dk_x dk_y A \left\{ \frac{1}{W} \left[ \mathbf{M}(\pm{k_{2z}}) + R^{TE}_1 \mathbf{M}({\mp}k_{2z}) \right] \right. \\
&\hspace{-8em}\left. \left[ R^{TE}_2 e^{-j2\phi_2} \mathbf{M'}({\pm}k_{2z}) + R^{TE}_1 \mathbf{M'}({\mp}k_{2z}) \right] + \right. \\
&\hspace{-8em}\left. \frac{1}{W'} \left[ \mathbf{N}({\pm}k_{2z}) + R^{TM}_1 \mathbf{N}({\mp}k_{2z}) \right]  \right. \\
&\hspace{-8em}\left. \left[ R^{TM}_2 e^{-j2\phi_2} \mathbf{N'}({\pm}k_{2z}) + R^{TM}_1 \mathbf{N'}({\mp}k_{2z}) \right] \right\}, \\
 \\
&\hspace{-12em}G^{e}_{sub}(\mathbf{R},\mathbf{R'}) = & \\
&\hspace{-10em}\iint_{0}^{\infty} dk_x dk_y A \\
&\hspace{-10em} \left\{ \mathbf{M}(-k_{3z})\frac{T^{TE}_3 }{W} e^{-j2(\phi_2-\phi_1)}\left[ \mathbf{M'}(k_{2z}) + R^{TE}_3 \mathbf{M'}(-k_{2z}) \right] \right. \\
&\hspace{-12em}\left. + \mathbf{N}(-k_{3z})\frac{T^{TM}_3 }{W'} e^{-j2(\phi_2-\phi_1)}\frac{k_{si}}{k_{sub}}\left[ \mathbf{N'}(k_{2z}) + R^{TM}_3 \mathbf{N'}(-k_{2z}) \right] \right\},
\end{aligned}
\label{eq:GreenFunctions1}
\end{equation}

\begin{backmatter}
\bmsection{Funding}
The author(s) declare that financial support was received for the research, authorship, and/or publication of this article. This work was made possible with the support of the NYUAD Research
Enhancement Fund.

\bmsection{Acknowledgments}
The authors express their gratitude to the NYU Abu Dhabi
Center for Smart Engineering Materials and the Center for Cyber
Security for their valuable contributions and support. Simulations for this research were partially carried out on the High-Performance Computing resources at NYUAD 

%\bmsection{Disclosures}
%``The authors declare no conflicts of interest.''

%\bmsection{Data availability} Data underlying the results presented in this paper are not publicly available at this time but may be obtained from the authors upon reasonable request.

\bigskip
\end{backmatter}

%%%%%%%%%% If using BibTeX:
\bibliography{sample}

%%%%%%%%%% If preparing manually:
% \begin{thebibliography}{1}
% \newcommand{\enquote}[1]{``#1''}

% \bibitem{Zhang:14}
% Y.~Zhang, S.~Qiao, L.~Sun, Q.~W. Shi, W.~Huang, L.~Li, and Z.~Yang,
%   \enquote{Photoinduced active terahertz metamaterials with nanostructured
%   vanadium dioxide film deposited by sol-gel method,}
%   {\protect\JournalTitle{Optics Express}} \textbf{22}, 11070--11078 (2014).

% \bibitem{Optica}
% {Optica}, \enquote{{Optica Publishing Group},}
%   \url{http://www.opg.optica.org}.

% \bibitem{FORSTER2007}
% P.~Forster, V.~Ramaswamy, P.~Artaxo, T.~Bernsten, R.~Betts, D.~Fahey,
%   J.~Haywood, J.~Lean, D.~Lowe, G.~Myhre, J.~Nganga, R.~Prinn, G.~Raga,
%   M.~Schulz, and R.~V. Dorland, \enquote{Changes in atmospheric consituents and
%   in radiative forcing,} in \enquote{Climate Change 2007: The Physical Science
%   Basis. Contribution of Working Group 1 to the Fourth assesment report of
%   Intergovernmental Panel on Climate Change,}  S.~Solomon, D.~Qin, M.~Manning,
%   Z.~Chen, M.~Marquis, K.~B. Averyt, M.~Tignor, and H.~L. Miler, eds.
%   (Cambridge University Press, 2007).

% \end{thebibliography}

\end{document}